\documentclass[12pt]{article}
\usepackage{graphics}
\usepackage{amssymb}
\usepackage{amsmath}

\newcommand{\rf}[1]{(\ref{#1})}
\newcommand{\beq}{\begin{equation}}
\newcommand{\eeq}{\end{equation}}
\newcommand{\bea}{\begin{eqnarray}}
\newcommand{\eea}{\end{eqnarray}}

\newcommand{\e}{\mbox{e}}
\renewcommand{\d}{\mbox{d}}
\newcommand{\g}{\gamma}
\newcommand{\G}{\Gamma}

\renewcommand{\b}{\beta}
\renewcommand{\a}{\alpha}

\newcommand{\m}{\mu}

\newcommand{\ep}{\varepsilon}

\newcommand{\om}{\omega}
\newcommand{\del}{\delta}
\newcommand{\Del}{\Delta}
\newcommand{\sg}{\sigma}

\newcommand{\oh}{\frac{1}{2}}

\newcommand{\ra}{\rangle}
\newcommand{\la}{\langle}
\newcommand{\prt}{\partial}
\newcommand{\mi}{\!-\!}

\newcommand{\cD}{{\cal D}}

\newcommand{\cM}{{\cal M}}

\newcommand{\cN}{{\cal N}}

\newcommand{\hg}{{\hat{g}}}
\newcommand{\hR}{{\hat{R}}}

\begin{document}

\begin{center}
\vspace{24pt}
{ \large \bf Roaming moduli space using dynamical triangulations}

\vspace{30pt}

{\sl J. Ambj\o rn}$\,^{a}$,
{\sl J. Barkley}$\,^{a}$,
and {\sl T.G. Budd}$\,^{b}$

\vspace{48pt}
{\footnotesize

$^a$~The Niels Bohr Institute, Copenhagen University\\
Blegdamsvej 17, DK-2100 Copenhagen \O , Denmark.\\
{ email: ambjorn@nbi.dk, barkley@nbi.dk}\\

\vspace{10pt}

$^b$~Institute for Theoretical Physics, Utrecht University, \\
Leuvenlaan 4, NL-3584 CE Utrecht, The Netherlands.\\
email: t.g.budd@uu.nl\\

}
\vspace{36pt}
\end{center}


\begin{center}
{\bf Abstract}
\end{center}
\noindent
In critical as well as in non-critical string theory
the partition function reduces to an integral 
over moduli space after integration over matter fields.
For non-critical string theory this moduli integrand
is known for genus one surfaces.
The formalism of dynamical triangulations  provides
us with a regularization of non-critical string theory.
We show how to assign in a simple and geometrical  way  
a moduli parameter to each triangulation. After integrating 
over possible matter fields we can thus construct the
moduli integrand. We show  numerically for $c=0$ and $c=-2$ 
non-critical strings that the moduli integrand converges
to the known continuum expression when the number 
of triangles goes to infinity.

\vspace{12pt}
\noindent

\vspace{24pt}
\noindent
PACS: 04.60.Ds, 04.60.Kz, 04.06.Nc, 04.62.+v.\\
Keywords: quantum gravity, lower dimensional models, lattice models.

\newpage

\section{Introduction}\label{intro}

Non-critical string theory, or two-dimensional Euclidean quantum gravity
coupled to matter, has been a fruitful laboratory for studying aspects 
of string theory as well as quantum gravity. Solving these theories  
one has had the advantage to have both a lattice version of the theory
and a continuum field theory formulation. The lattice version has 
been denoted the dynamical triangulation model ($DT$) or the matrix model of 
non-critical string theory.  It can be solved, basically by combinatorial 
methods, exemplified by the use of matrix models. 
The explicit solution of the continuum model uses the fact that 
the Liouville theory is a (special) conformal theory. 
For observables which can be calculated by both approaches agreement is
found. 

Since  non-critical string theory is described by conformal 
field theory we know that conformal invariance is implemented. However, the 
precise manifestation of this invariance, and how 
it is related to the moduli space of the  
underlying surfaces (i.e.\ to  the  part 
of the surface geometry left invariant under conformal 
transformations) has been rather limited. That dynamical 
triangulations or matrix models contain precise information
about the moduli space is on the other hand obvious. For instance it was shown 
in \cite{ackm} that the resolvent of the matrix model  
has an expansion in terms of so-called 
moments, where the coefficients in the double scaling
for the genus $h$ terms are precisely the intersection indices of
moduli space for genus $h$ Riemann surfaces with any 
number of punctures. It was also shown
how these matrix models in the double scaling limit can be 
related to the Kontsevich matrix model which provides a representation
of the generating function for these intersection indices \cite{ackm,ak}. 
Much work has later expanded on these
results \cite{bertrand}, but we are in these approaches far from the naive and 
simple idea which was the starting point for $DT$ and 
matrix models, namely that $DT$ provides a  regularization
of the path integral of non-critical strings and thus the 
moduli parameters should appear in the same simple way as 
they formally appear in the string path integral.    

One problem when using matrix models is that most observables
which can be calculated analytically are of global nature: integrated 
correlation functions of matter fields or the partition function 
as a function of various boundary states. Further, the comparison
with continuum results is only possible for the simplest topologies
of the surfaces: spherical topology, disc topology
and (only recently \cite{cylinder}) cylindrical topology.
It is simply difficult to perform 
calculations in the framework of Liouville theory for higher genus 
surfaces. However, there is a narrow window where one can test
in more detail if the matrix models, 
or the framework of dynamical triangulations,
actually agree with the continuum expressions for higher genus surfaces
in the naive way mentioned above.
We  know the partition function for genus one
surfaces of non-critical strings expressed as an 
integral over the moduli parameter of the torus \cite{genus1}. Integrating out
the matter fields, using the conformal anomaly, worldsheet conformal
invariance ensures that the remaining integrand of the partition function is 
only an (explicitly known) function of the moduli parameters. We can 
now in principle compare this integrand to the integrand which we obtain
using dynamical triangulations as a regularization of the non-critical string
theory. Which integrand {\it do} we obtain using dynamical triangulations?
For each triangulation we have a continuous, piecewise linear geometry.
To each such geometry we can associate moduli parameters (as described below).
Each geometry will appear with a certain weight which depends on the 
matter we have coupled to the surface. If we have $N$ triangles we will
in this way obtain a number of {\it points} in moduli space 
and when $N$ goes to 
infinity we expect these discrete points converge to a density distribution 
which should be proportional to the integrand of non-critical string theory. 
The possibility of performing such a comparison 
was pioneered by Kawai and collaborators\cite{kawaigenus}.
They found good qualitative agreement 
between the non-critical string integrand and the density 
constructed from dynamical triangulations. The purpose of this article is 
to improve this test, making it quantitative, and also 
present a more general setup than the one used in \cite{kawaigenus}.
We also show how one can measure the moduli parameters for 
higher genus triangulations, but unfortunately there is presently 
no theoretical calculation with which we could compare such results
which would be easy to generate numerically.

\section{The setup}\label{setup}

In string theory we can integrate out the conformal matter fields
on the worldsheet by the conformal anomaly. This is true also for 
non-critical string theory. 
The partition function of e.g.\ a bosonic string in $d$ dimensions
can be written as 
\beq\label{2.0}
Z^{(h)} = \int \d \m(\tau_i) \; \int \cD_{\hg} \phi\, \cD_\hg b\, \cD_\hg c\, 
\cD_\hg X_\m \;\e^{-S(X,\hg)-S(b,c,\hg)-S_L(\phi,\hg)}, 
\eeq
where the integration over worldsheet geometries 
of genus $h$ is implemented by integrating over
metrics $g_{\a\b}$, gauge fixed to conformal gauge 
$g_{\a\b}= e^\phi \hg_{\a\b}(\tau_i)$, $\hg_{\a\b}(\tau_i)$ denoting
a fiducial background metric depending on (for $h>1$) $3h\mi 3$ complex
moduli parameters $\tau_i$ parametrizing the complex structure of the
genus $h$ Riemann surface and $\d \m(\tau_i)$ 
being a modular invariant measure.
The $b,c$ denote the Faddeev-Popov ghosts associated with the partial gauge
fixing to conformal gauge, $\phi$ denotes the Liouville field  and finally 
$S_L(\phi,\hg)$ is the Liouville action:
\beq\label{1s} 
S_L(\phi,\hg) = \frac{1}{4\pi} \int d^2 \xi \sqrt{\hg(\xi)} \; 
\Big( (\prt_\a \phi)^2 + Q\, \hR\,\phi +\mu \,\e^{2\b \phi} \Big)
\eeq
The constants in the 
Liouville action are chosen such that the conformal invariance 
is maintained in the quantum theory
in accordance with the bootstrap approach of David, Distler and Kawai
\cite{DDK}, i.e.\
\beq\label{2s}
Q = \sqrt{(25-d)/6},~~~~Q = 1/\b + \b.
\eeq

For a fixed value of the moduli parameters $\tau_i$ one can in 
principle perform the integration over the bosonic fields and 
the ghost fields. One obtains contributions 
from zero modes and appropriate determinants. However, only 
in the case of genus one is the integration over the zero
mode trivial, the reason being that $\int d^2 \xi \sqrt{\hg(\xi)} \, \hR =0$.
In this case one obtains just the standard result from critical string 
theory \cite{genus1}. Thus, if we consider the situation
where we fix the world sheet area to be $A$, i.e.\ trade the 
cosmological constant $\m$ in \rf{1s} for the area $A$, we obtain
\beq\label{3s}
Z^{(h=1)}(A) \sim A^{-1}\int_\cM \frac{d^2\tau}{\tau_2^2}\; F(\tau)^{c-1}.
\eeq    
In \rf{3s} $c$ denotes the central charge of the conformal matter
theory coupled to 2d gravity, $\tau=\tau_1+i\tau_2$ the complex 
moduli parameter of the complex structure we are integrating over
and $\cM$ is the region in the complex plane to which $\tau$ belongs.
Finally $F(\tau)$ is given by: 
\beq\label{4s}
F(\tau) = \tau_2^{-1/2} \e^{\pi \tau_2/6} \prod_{n=1}^{\infty} 
|1-\e^{2\pi i n \tau}|^{-2}.
\eeq

We want to test how well the non-critical string theory regularized 
by $DT$ approximates this formula. We will further restrict ourselves to the 
cases $c=0$ and $c=-2$. We do so in order to obtain
the best statistics in the numerical tests 
we perform. The case $c=0$ is chosen to avoid to have to update matter fields
in addition to the geometry, when we use Markov chain Monte Carlo simulations
to generate configurations. The case $c=-2$ is chosen in order to
have a non-trivial matter system coupled to gravity which can 
be easily handled numerically. For $c=-2$ there exists a 
recursive algorithm which allows us to avoid constructing Markov chains
and generate directly random triangulations with the correct weight,
including the matter fields \cite{kk}.

The set-up is thus the following: 
Using $DT$ we can fix the area, like in \rf{3s}.
This is done by fixing the number of triangles, since all triangles
are viewed as identical in the $DT$ formalism. A triangulation can thus in 
principle be viewed as representing a piecewise linear geometry.
As we will show below there is a natural way to associate
with such a piecewise linear geometry a moduli parameter.
For the ensemble of piecewise linear geometries constructed from
$N$ identical equilateral triangles, each triangulation 
with the topology of a torus, we will then have an approximation
to the continuum distribution $\tau_2^{-2}F(\tau)^{c-1}$. We will investigate
if and how the corresponding distribution of the moduli parameter
$\tau$ constructed from the $DT$ ensemble characterized by $N$ triangles
converges to the continuum distribution for $N \to \infty$. 
A priori it is not clear that there should be a convergence at all. Given 
a continuum area $A$, we usually think of the lattice area as 
$A = N a^2 \sqrt{3}/4$, $a$ being the link length of the 
triangulation. Thus keeping $A$ fixed and taking $N \to \infty$
dictates how one should think of the lattice link length $a \to 0$.
However, the corresponding piecewise linear surface will in
general not be anything like a smooth surface. In fact we know
now that it with probability one will be fractal with a 
Hausdorff dimension different from two \cite{fractal}. The situation 
is thus quite similar to the one encountered for the ordinary 
path integral of a particle: with probability one a path in the 
path integral is nowhere differentiable and has a Hausdorff dimension 
different from one. As long as we think of the $DT$ ensemble of 
piecewise linear surfaces in this way one would not be too worried
about the highly fractal structure of the geometry associated 
with a generic triangulation in the ensemble for $N\to \infty$,
and we know that for global observables one obtains
identical results using continuum Liouville theory and 
the $DT$ ensemble, taking $N\to \infty$. However, it is less 
clear if an assignment of a moduli parameter as we are going
to do it, by analogy between differential forms and ``discrete 
differential forms'' on simplicial complexes, even if very natural, 
will work for the  ``wild'' generic triangulations we meet in the 
$DT$ ensemble. Recall for instance that the spectral properties of 
the Laplacian defined even on the simplest piecewise linear surfaces
are different from the generic spectral properties of the Laplacian 
defined on a smooth geometry, because one should, when analyzing 
the spectrum, view the piecewise linear geometry as a geometry
with conical singularities at the vertices \cite{conical}.
Since we are actually using properties of the Laplacian to determine
the moduli parameter, one could be worried whether the assignment using 
discrete differential forms makes sense in the present context.   
However, we will find that the assignment works beautifully!

Let us start by showing how to determine the moduli parameter $\tau$ if 
given a two-dimensional Riemannian geometry of a genus one surface. 
Let $d$ denote the exterior derivative and $\del$ its adjoint 
(the co-differential) with respect to the standard inner product
defined on $p$-forms, $p=0,1,2$ for two-dimensional surfaces:
\beq\label{5s}
\la \phi_p | \psi_p\ra = \int  \phi_p \wedge *\psi_p.
\eeq
where $*\psi_p$ is the Hodge dual\footnote{In components the Hodge dual
of a $p$-form $\psi_p$ on a $n$-dimensional Riemannian manifold is defined as:
\bea
\psi_p &=& \frac{1}{p!} \psi_{a_1\ldots a_p} \,dx^{a_1}\wedge \cdots \wedge
dx^{a_p},\nonumber\\
*\psi_p &=&\frac{\sqrt{g}}{p!(n-p)!} \,
\ep_{a_1\ldots a_p,b_1\ldots b_{n-p}}
\psi^{a_1\ldots a_p}\, dx^{b_1}\wedge \cdots \wedge dx^{b_{n-p}} 
\nonumber
\eea
} $2\mi p$ form of $\psi_p$. In particular 
we have for two 1-forms:
\beq\label{6s}
\la \phi_1|\psi_1\ra = \int d^2x \sqrt{g}\, g^{ab}\phi_a \psi_b,
\eeq
where $\phi_1 = \psi_a(x)dx^a$ and $\psi_1 = \psi_a(x)dx^a$, $a=1,2$ 
in some coordinate system $x^a$.
The Hodge Laplacian 
\beq\label{7s}
\Del = d \del +\del d 
\eeq
maps $p$-forms to $p$-forms, and for 1-forms on a genus $h$ surface
the kernel is $2h$-dimensional, constituting 
the so-called harmonic differentials. For a genus 
one surface the vector space of harmonic 1-forms is thus 
two-dimensional. We can find it by solving 
\beq\label{20s}
d \phi_1 = 0,~~~~~~\del \phi_1 =0,
\eeq
which express that a harmonic 1-form has a vanishing curl and divergence.

Let us choose two closed curves $\g_1$ and $\g_2$
on the torus that generate the fundamental group. It is 
then possible to choose ``dual'' harmonic 1-forms $\a^1$ and $\a^2$, i.e.\
harmonic forms which satisfy:
\beq\label{8s}
\int_{\g_i} \a^j = \del_i^j.
\eeq
 
For the torus the moduli parameter can be defined in the following way:
by the uniformization theorem any metric on a surface is conformally 
related to a constant-curvature metric. For the torus it is
a zero-curvature metric, and starting from the Euclidean plane 
we obtain such a flat torus by identifying opposite sides in a parallelogram.
Let the  two sides of the parallelogram be vectors $\om_1$
and $\om_2$, represented as complex numbers. The moduli parameter $\tau$
is defined as $\tau= \om_2/\om_1$. It is clearly invariant under 
translations, rotations and global scaling of the parallelogram and we can assume that $\tau_2 >0$. However, there are finite 
diffeomorphisms of the torus which cannot 
be obtained continuously from the identity, and which 
we will insist do not change the partition function. These transformations
do not leave $\tau$ invariant but will change it according to 
\beq\label{30s}
\tau \to \frac{a \tau +b}{c\tau +d},
~~~~~~~~ \begin{matrix} a,b,c,d \in \mathbb{Z} \\
 ad-bc = 1 \end{matrix}.
\eeq
The transformations \rf{30s} constitute a transformation group 
called the modular group $\G$ and it is isomorphic
to $SL(2,\mathbb{Z})/\mathbb{Z}_2$. In the integral \rf{3s} we are instructed 
only to integrate over points $\tau$ which cannot be identified by 
the action of the modular group via \rf{30s}, i.e.\ $\cM$ is 
a so-called fundamental domain of $\G$ in the upper half-plane:
no pair of points within it can be connected by a modular transformation
and any point outside it can be reached from a unique point inside by 
some modular transformation. A standard choice of fundamental domain $\cM$ is:
\beq\label{31s}
\tau \in \cM ~~{\rm if}~~~
\left\{\begin{array}{lll} \tau_2 >0,~~-\oh \leq \tau_1 \leq 0 & ~~{\rm and} & 
~~|\tau|\geq 1 \\
 \tau_2 >0,~~~~~~ 0 < \tau_1 < \oh & ~~{\rm and} & 
~~|\tau|> 1 
\end{array}\right.
\eeq
We will use this fundamental domain in the rest of the article. Of course
it is not unique: for example any modular transformation will transform it to another
fundamental domain.

Given a metric $g_{ab}$ and the curves $\g_i$, 
$\tau$ is determined by the formula
\beq\label{9s}
\tau = - \frac{\la \a^1 | \a^2\ra}{ \la \a^2 | \a^2\ra}
+i \sqrt{\frac{\la \a^1 | \a^1\ra}{ \la \a^2 | \a^2\ra} -
\left(\frac{\la \a^1 | \a^2\ra}{ \la \a^2 | \a^2\ra}\right)^2}.
\eeq

To prove this formula we note that we can choose periodic coordinates $x^1$ and $x^2$, both with period 1, such that $\g_1$ and $\g_2$ run around in the $x^1$ and $x^2$ direction,
respectively and such that metric takes the form:
\beq\label{50s}
ds^2 = \e^{\sg (x)} \hg_{ab} dx^a dx^b,
\eeq
where $\e^{\sg(x)}$ is a local scale factor and 
$\hg$ represents a flat metric with determinant 1.
The parallelogram can now be represented in the Euclidean plane
as shown in Fig.\ \ref{torus} and the $\hg$ is thus explicitly 
given by 
\beq\label{51s}
\hg_{ab} = \frac{1}{\tau_2} \begin{pmatrix} 1 & \tau_1 \\ \tau_1 & 
\tau_1^2 + \tau_2^2 \end{pmatrix},
\eeq 
the factor $1/\tau_2$ enforcing $\det \hg_{ab} =1$.
Since harmonicity of the 1-forms is preserved under conformal transformations,
we clearly have that $\a^1 = dx^1$ and $\a^2=dx^2$ and the inner product 
$\la \a^i | \a^j\ra $ can be read off  from the definition \rf{6s}:
\beq\label{52s}
\la \a^i | \a^j\ra = \hg^{ij},~~~~{\rm i.e.}~~~~
\la \a^1 | \a^1\ra \la \a^2 | \a^2\ra-\la \a^1 | \a^2\ra^2=1.
\eeq 
Eq.\ \rf{9s} follows from \rf{51s} and \rf{52s}.

The $\tau$ determined
by  formula \rf{9s} lies in the upper half-plane, but 
not necessarily in the fundamental domain $\cM$ defined in \rf{31s}. 
We use \rf{30s} to find the modular transformation which maps 
it to $\cM$.

\begin{figure}[t]
\centerline{\scalebox{0.8}{\rotatebox{0}{\includegraphics{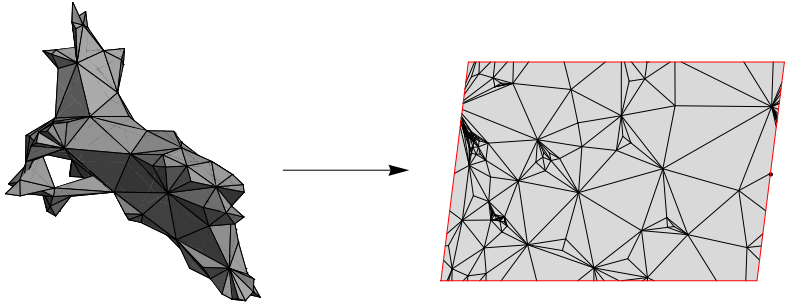}}}}
\caption{The embedding of a triangulation of the torus into the 
Euclidean plane. This particular triangulation with $N=280$ triangles 
is taken from the ensemble $DT$(3) and has moduli parameter 
$\tau = 0.09 + 0.7 i$. } 
\label{reconstruction}
\end{figure}

The formula \rf{9s} can be applied directly to the piecewise linear
geometries encountered in the $DT$ formalism since the notion of $p$-forms
and an exterior derivative acting on these forms can be defined in a natural 
way. For a detailed definition for piecewise linear geometries, defined
by  triangulations  where the lengths $\ell_{ij}$ of the links between 
vertices $i$ and $j$ are given, we refer to standard textbooks \cite{ID}.
In the case of $DT$ the formulas become very simple because all link
lengths are identical, and similarly all areas of triangles are identical.
Thus for the $DT$ piecewise linear surfaces the definitions
become identical to the ``abstract'' definition of 
discrete differentials used on simplicial complexes to
study the discrete versions of cohomology as a more general 
setup than the De Rham cohomology of ordinary differential forms.
 A discrete $p$-form $\phi_p$ is in the simplicial complex  context
defined by assigning a real number 
$\phi_p(\sg_p)$ to each 
oriented $p$-simplex $\sg_p$. Changing orientation of the simplex changes 
the sign of the assigned real number. The linear space $\Omega_p$ of 
$p$-forms has dimension equal to the number of $p$-simplices in the 
triangulation. For the scalar product corresponding to \rf{5s} we  
simply choose
\beq\label{10s}
\la \phi_p | \psi_p \ra = \sum_{\sg_p} \phi_p(\sg_p) \psi_p(\sg_p).
\eeq
The exterior derivative is defined as 
\beq\label{11s}
(d\phi)_{p+1}(\sg_{p+1}) = \sum_{\sg_p \in \sg_{p+1}} (-1)^{\sg_p} 
\phi_p(\sg_p),
\eeq
where the sum is over all $p$-subsimplices in $\sg_{p+1}$ and the sign
factor signifies the orientation of the subsimplex relative to the orientation
of $\sg_{p+1}$. Again $\del$, the co-differential, is defined as the 
adjoint of $d$ with respect to the scalar product \rf{10s}. To be
explicit, consider the 1-forms. The 1-forms take values on the oriented links.
Denote the vertices in the triangulation $i$ and the links $(ij)$ when
vertices $i$ and vertices $j$ are connected by a link oriented 
from $i$ to $j$. Similarly
triangles are denoted $(ijk)$. Thus we have
$\phi_{(ij)} = -\phi_{(ji)}$ and 
\beq\label{12s}
(d\phi)_{(ijk)} = \phi_{(ij)}+\phi_{(jk)}+\phi_{(ki)},~~~(\del \phi)_i = 
\sum_{j(i)} \phi_{(ij)},
\eeq
where the summation $j(i)$ is over all links incident on vertex $i$.
If one wants to relate 1-forms defined this way to continuum 
notation, one can heuristically think of the 1-form $\phi_{(ij)}$ as 
\beq\label{13s}
\phi_{(ij)}= \int_i^j dx^a \, \phi_a(x).
\eeq
Using \rf{13s} and appealing to Stokes theorem $(d\phi)_{(ijk)}$ will be 
proportional to  the curl 
associated with triangle $(ijk)$, and similarly Gauss' theorem 
makes $(\del\phi)_i$ proportional to the 
divergence assigned to the cell dual to vertex $i$. We note
that using the definition \rf{12s} it is easy to 
show  \cite{ID} that for a triangulation of genus $h$ the 
equations  $(d\phi)_{(ijk)} =0$ and 
$(\del \phi)_i=0$ have $2h$ solutions, in agreement 
with the continuum result referred to above.
      
Finally we define a path $\g$ 
in the triangulation as a path of  links.
Thus we have a natural integration of 1-forms along a path:
\beq\label{14s}
\int_\g \phi = \sum_{(ij)\in \g} \phi_{(ij)}.
\eeq

We now have all the ingredients for the construction of the moduli parameter
$\tau$ for a piecewise linear triangulation belonging to the $DT$ ensemble
of genus 1. First identify two non-contractible loops $\g_1$ and $\g_2$ which 
generate the fundamental group. Next solve $(d\phi)_{(ijk)} =0$ and
$(\del \phi)_i=0$ to find the two-dimensional space of harmonic 1-forms. 
Finally select from this space the two 1-forms 
$\a^1$ and $\a^2$ which satisfy the discrete analogy of \rf{8s}, i.e.\
\beq\label{15s}
\sum_{(kl) \in \g_i} \a^{j}_{(kl)} = \del^j_i,
\eeq
and use \rf{9s} (and \rf{10s}) to calculate $\tau$.           

Let us finally note that we can use the harmonic 1-forms $\a^1$ and
$\a^2$ to obtain an explicit embedding of a $DT$ triangulation into 
a parallelogram in the Euclidean plane. In the continuum,
choosing coordinates such that the generating curves $\g_1$ and 
$\g_2$ run around in the $x^1$ and $x^2$ directions, respectively,
we have $\a^i =dx^i$. Thus, knowing $\a^i$ we can recover the 
coordinates $x^i$ by integration. We now apply this to our 
triangulation using the two harmonic 1-forms $\a^{i}_{(kl)}$
we have constructed for the triangulation. For each link 
$(kl)$ we view the vector $(\a^1_{(kl)},\a^2_{(kl)})$ as representing the 
link $(kl)$ in the Euclidean plane precisely in
the heuristic sense of eq.\ \rf{13s}. Starting from an arbitrary vertex 
we can now reconstruct the triangulation in the Euclidean plane.
Fig.\ \ref{reconstruction} shows an example of such a reconstruction.
Recall that a  function is harmonic if and only if it 
has the mean value property: the value of a harmonic function at 
the center of a circle is equal to the integral of the function 
along the circle divided by the length of the circle.  
The discrete map from the triangulation to the Euclidean plane 
is harmonic in the sense that any vertex is located at the center 
of mass of its neighbors. 

\begin{figure}[t]
\centerline{\scalebox{0.65}{\rotatebox{0}{\includegraphics{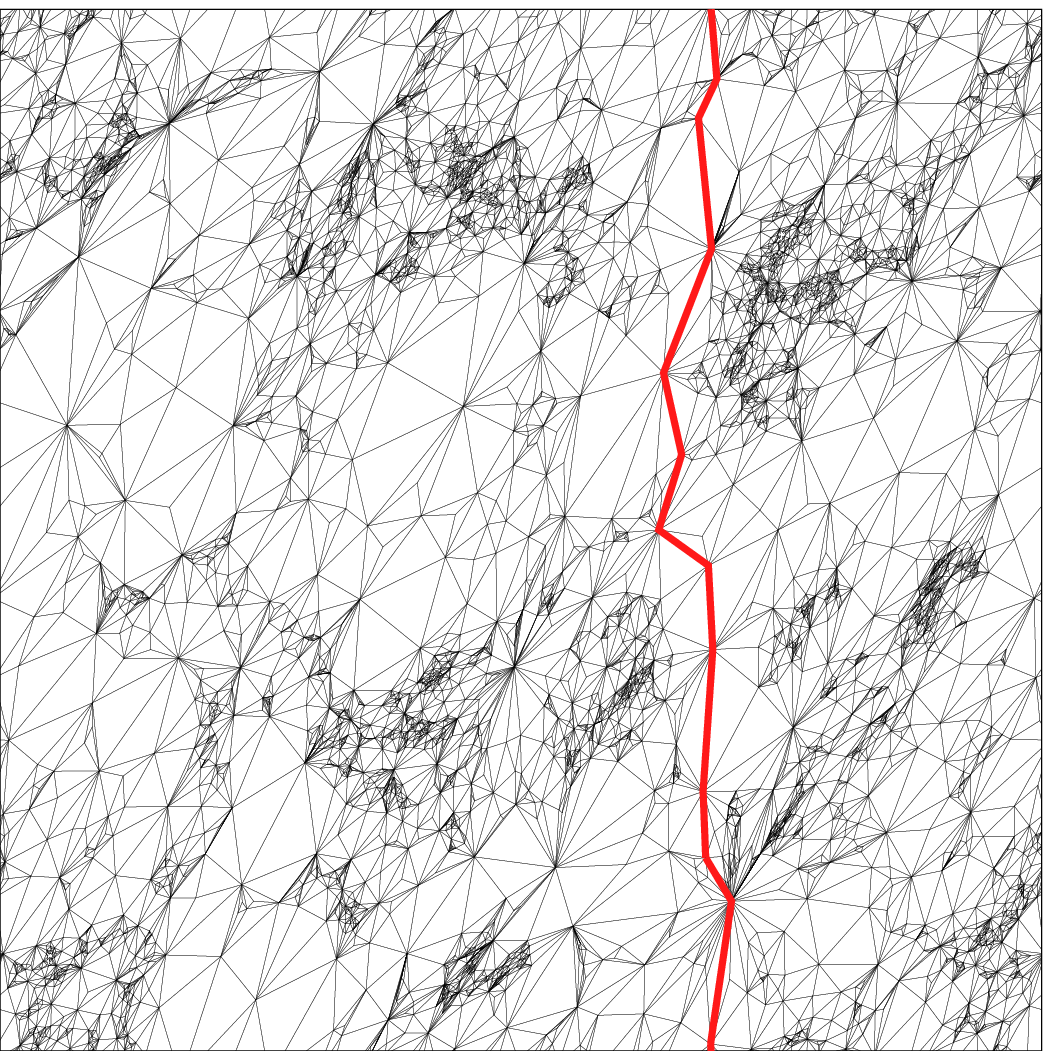}}}
~~~~\scalebox{0.65}{\rotatebox{0}{\includegraphics{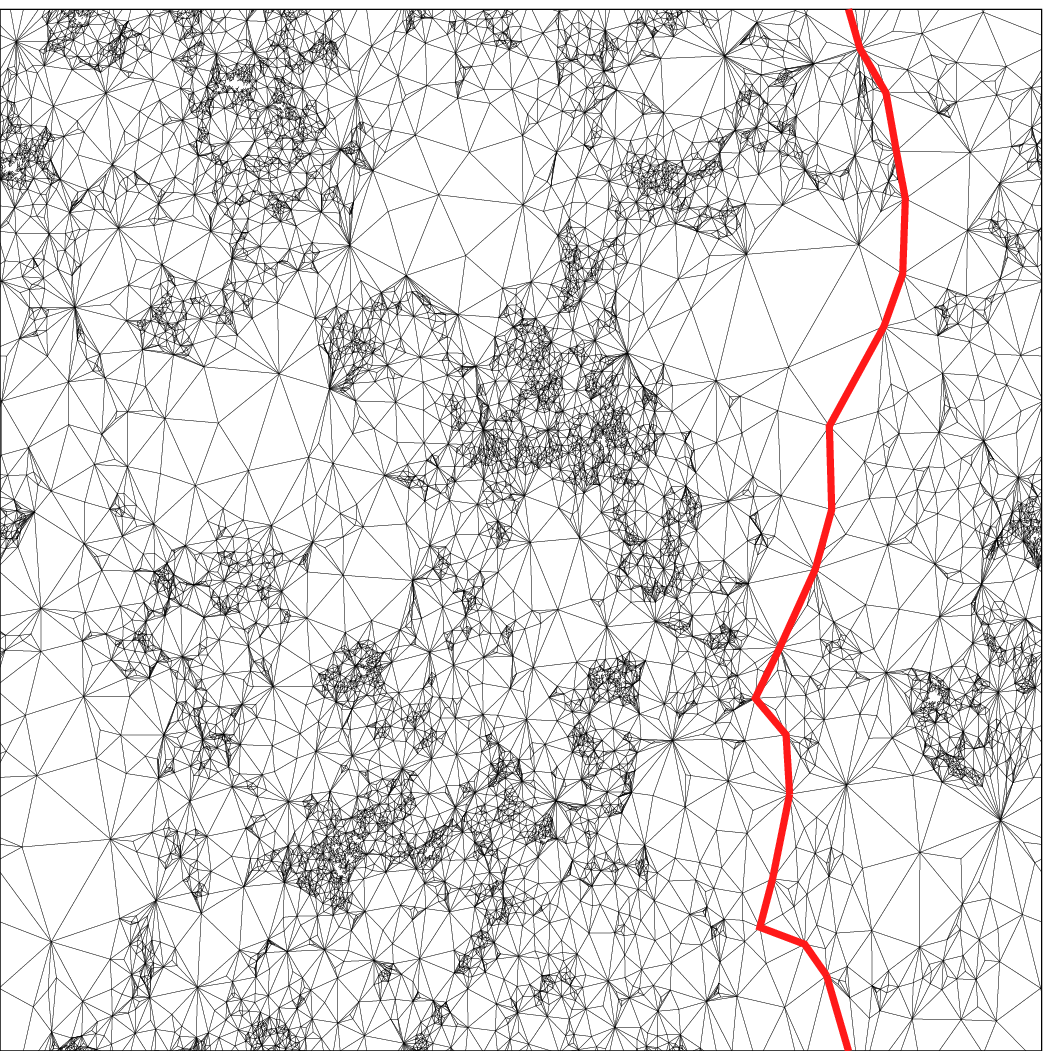}}}}
\caption{Embeddings of two large triangulations of the torus into the 
Euclidean plane. The triangulations have $150000$ triangles.
The triangulation to the left is picked from the ensemble of $c=0$ 
triangulations (2d gravity without matter fields), 
while the triangulation to the right is from 
the $c=-2$ ensemble (to be described in Sec. \ref{c-2a}).  
Shown in red are the shortest non-contractible loops.}
\label{reconstructionc-2}
\end{figure}

Fig.\ \ref{reconstructionc-2} shows
other embeddings of  triangulations of the torus in the Euclidean plane.
These triangulations are much larger than the one shown in 
Fig.\ \ref{reconstruction}, each consisting  of $N=150000$ triangles.
The triangulation to the left is from the $c=0$ ensemble 
of triangulations considered in the next section 
(2d gravity without matter fields), while the 
triangulation to the right is picked from the 
ensemble of triangulations describing conformal $c=-2$ matter 
coupled to 2d gravity (to be described in detail in Sec.\ \ref{c-2a}).
Shown in red are also the shortest non-contractible loops, which we will discuss later.
In addition Fig.\ \ref{reconstructionc-2} captures qualitatively 
the fractal structure of the ensemble of triangulations, to be 
analyzed below. 

The method described above for calculating  the moduli 
of a genus one surface (and the moduli of a genus one triangulation)
has a relatively simple generalization to 
higher genus surfaces and triangulations. This is outlined in appendix \ref{highergenus}.
Below we describe the result of computer simulations which extract 
the moduli of genus one triangulations. It is unproblematic
to do the same for  genus $h$ triangulations, $h>1$, although
the moduli space now has the (real) dimension $6h-6$, but we have no continuum
expressions with which we can compare the sampled distributions
of moduli parameters, as already mentioned. 

\section{Measuring $\boldsymbol{\tau}$ for $\boldsymbol{c=0}$}\label{c0}

\begin{figure}[t]
\centerline{\scalebox{0.3}{\rotatebox{0}{\includegraphics{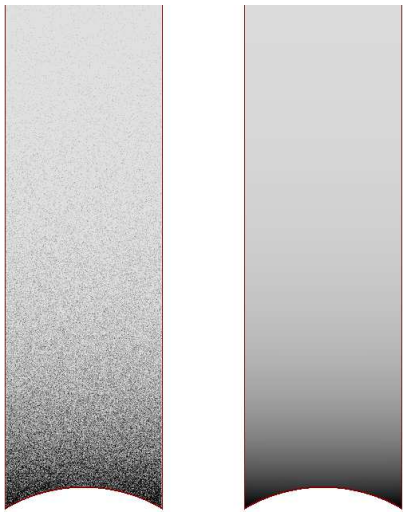}}}}
\caption[fig1]{(Part of) the fundamental domain. The left figure shows
the density obtained from (a small number of)
actual measurements for 1000 triangles, the right figure the 
theoretically calculated density.} 
\label{fundamental}
\end{figure}

In order to measure $\tau$ for $c=0$ we perform Monte Carlo simulations
starting out with a triangulation of the torus, constructed from $N$
triangles. For a description of how to perform such simulations
we refer to \cite{book}. The Monte Carlo simulations consists
of ``moves'' which change the triangulation locally, preserving $N$ and
the topology. After sufficiently many local moves we will have 
created a statistically independent triangulation. The number of 
moves needed on average for this increases with $N$. From the
independent configurations constructed this way 
we now calculate $\tau$ by 
\begin{itemize}
\item[(1)]
identifying 
two non-contractible link-loops $\g_1$ and $\g_2$ which generate the fundamental group of the torus, 
\item[(2)] constructing the 
harmonic differentials $\phi_{(ij)}^{harm}$ 
by solving the divergence and curl equations,
$(d\phi)_i=0$ and $(\del \phi)_{ijk} =0$,
\item[(3)] forming   linear combinations such that \rf{15s} is satisfied,
\item[(4)] constructing $\tau$ using \rf{9s}, 
\item[(5)] finding  the modular transformation
\rf{30s} which maps $\tau$ to the fundamental domain $\cM$ defined
in eq.\ \rf{31s}. 
\end{itemize}
In this way we construct a distribution of $\tau$'s which we can compare 
with the theoretical distribution. The most computer intensive part
of the calculation is step (2), 
the identification of the two-dimensional subspace
of (discrete) harmonic 1-forms. Fig.\ \ref{fundamental} shows a plot of 
the $\tau$-distribution. 

\begin{figure}[t]
\centerline{\rotatebox{0}{\includegraphics{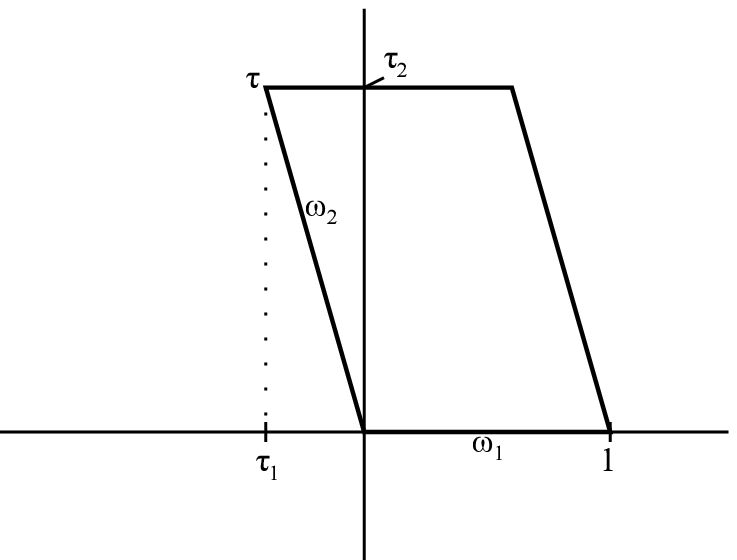}}}
\caption[fig1]{The parallelogram rescaled and placed such that 
$\om_1$ is the vector (1,0) and $\om_2=(\tau_1,\tau_2)$.} 
\label{torus}
\end{figure}

The fundamental domain $\cM$ is non-compact. If we place 
the flat torus as shown in Fig.\ \ref{torus}  we clearly have
a torus very elongated in the vertical direction for large 
imaginary values of $\tau$. 

\begin{figure}[t]
\centerline{\scalebox{1.1}{\rotatebox{0}{\includegraphics{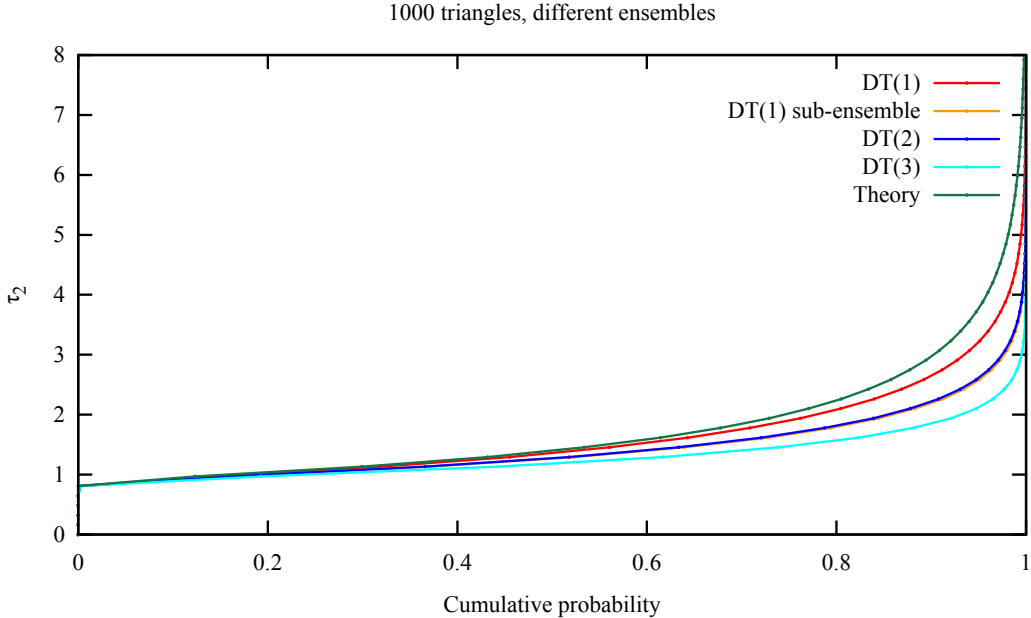}}}}
\caption[fig1]{The cumulative probability of the various $DT$-ensembles plotted
against $\tau_2$. On the plot it is difficult
to distinguish the curves for $DT(2)$ and the 
$DT(1)$ sub-ensemble with no non-contractible loops of length 1.
The $DT(1)$ curve is clearly closest to the theoretical curve.}
\label{DT(i)}
\end{figure}

Of course this does not necessarily reflect a property of the original
$DT$ triangulation since we are using a (discrete) conformal map
to embed the triangulation in the Euclidean plane. However, on 
average one would expect a large imaginary value of $\tau$ to 
reflect a $DT$-torus with   short non-contractible loops in one 
``direction''  compared to the possible lengths of non-contractible
loops in the other ``direction''. For the triangulations there are
two restrictions entering: by definition there is a shortest allowed loop
length since we count the loop length in an integer number of links. 
Also, for a finite $N$, there is a limit to how long  one can make
geodesic loops. Thus we expect for a finite $N$ that our 
$DT$-distribution of $\tau$'s will fail for large imaginary values of 
$\tau$, but it should improve  if (I) we increase
$N$ and (II) we use $DT$-ensembles which allow smaller non-contractible
loops. The second point can easily be tested in the following way: let 
us start with a $DT$-ensemble of ``regular'' triangulations $DT(3)$. By that 
we mean triangulations where each vertex has at least order three
and where two vertices are connected by at most one link. The dual 
graphs are $\phi^3$ graphs without tadpole and self-energy subgraphs. 
On this ensemble a shortest non-contractible loop is of length 3. 
We could instead allow two vertices to be connected by two different links,
belonging to different triangles. We call this ensemble $DT(2)$. The 
dual graphs are $\phi^3$ graphs without tadpoles, but with self-energy
subgraphs (except the simplest ones which would result in vertices of order
two). The shortest non-contractible loop length for this ensemble is 
clearly 2. Finally we consider the ensemble where we allow  a link to 
loop to its own vertex. In terms of dual graphs we now allow tadpole
graphs (but exclude still the simplest self-energy graphs to maintain
that all vertices in the triangulation have an order larger than or 
equal three\footnote{We should stress that there is nothing 
important related to requirement 
that the vertex order is larger than or equal 3. We just maintained 
it because our original computer program had it built in.}).
We denote this ensemble $DT(1)$. The shortest loop length
is one for this $DT$-ensemble. In Fig.\ \ref{DT(i)} we have shown the 
probability distribution of $\tau_2$ for a fixed $N$ and compared it
to the theoretical, continuum distribution. The $DT(1)$ distribution 
is significantly closer to the continuum distribution for large $\tau_2$
as expected. We also checked explicitly that the improvement was due
to non-contractible loops of length 1 by calculating the distribution
of $\tau_2$ for the sub-ensemble of $DT(1)$ which had no non-contractible 
loops of length 1. The results then agreed with the results for the 
$DT(2)$ ensemble.  In the rest of this article we will thus only 
use the ensemble $DT(1)$, which we for simplicity just call the $DT$-ensemble.

\begin{figure}[t]
\centerline{\rotatebox{0}{\includegraphics{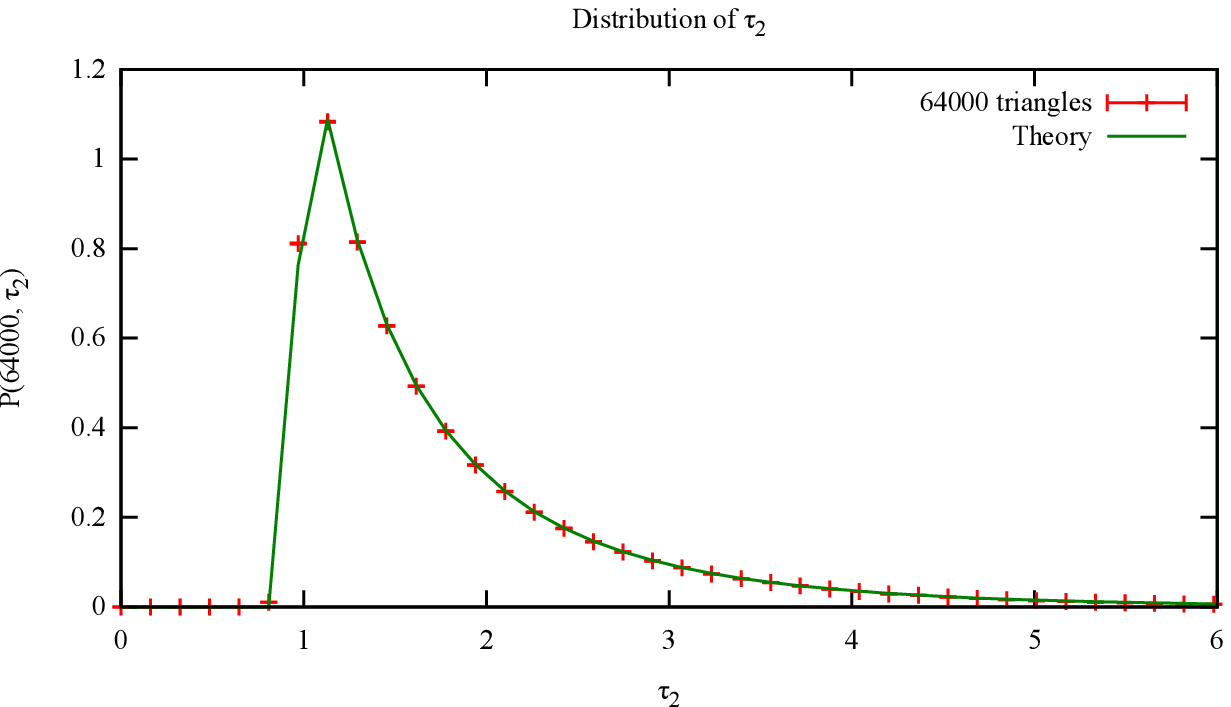}}}
\caption{The $\tau_2$ distribution for $N=64000$,
compared to the theoretical distribution.}
\label{tau2-distribution}
\end{figure}

\begin{figure}[t]
\centerline{\rotatebox{0}{\includegraphics{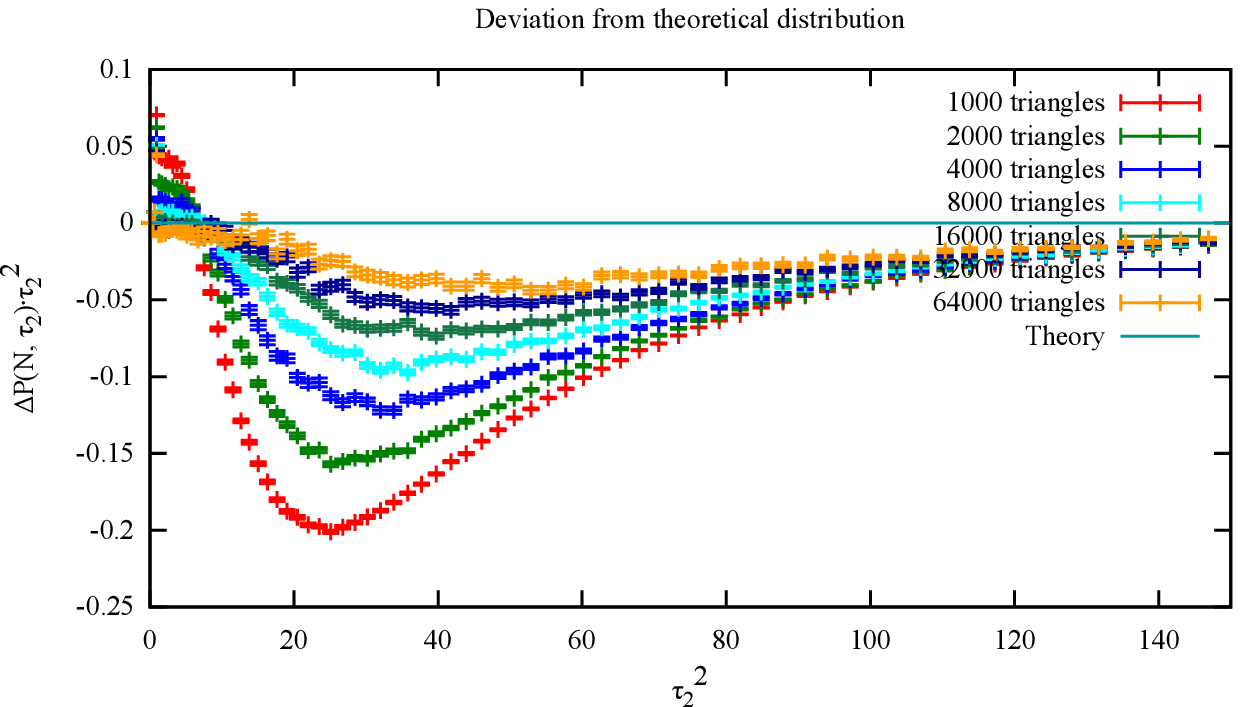}}}
\caption{$\tau_2^2 \Del P(N,\tau_2)$, 
$\Del P(N,\tau_2)\equiv P(N,\tau_2)-P({\rm theory},\tau_2)$, plotted as 
a function of $\tau_2^2$ for $N=2^K\cdot 10^3$, $K=0,1,2,\ldots 6$}
\label{N-dependence}
\end{figure}

Let us now turn to point (I) above, the $N$ dependence. Fig.\ 
\ref{tau2-distribution} shows that the $\tau_2$ distribution agrees
very well with the theoretical distribution for the large 
value of $N$ used, and Fig. \ref{N-dependence}
shows the deviation from the theoretical distribution for different $N$'s.

\begin{figure}[t]
\centerline{\scalebox{1.0}{\rotatebox{0}{\includegraphics{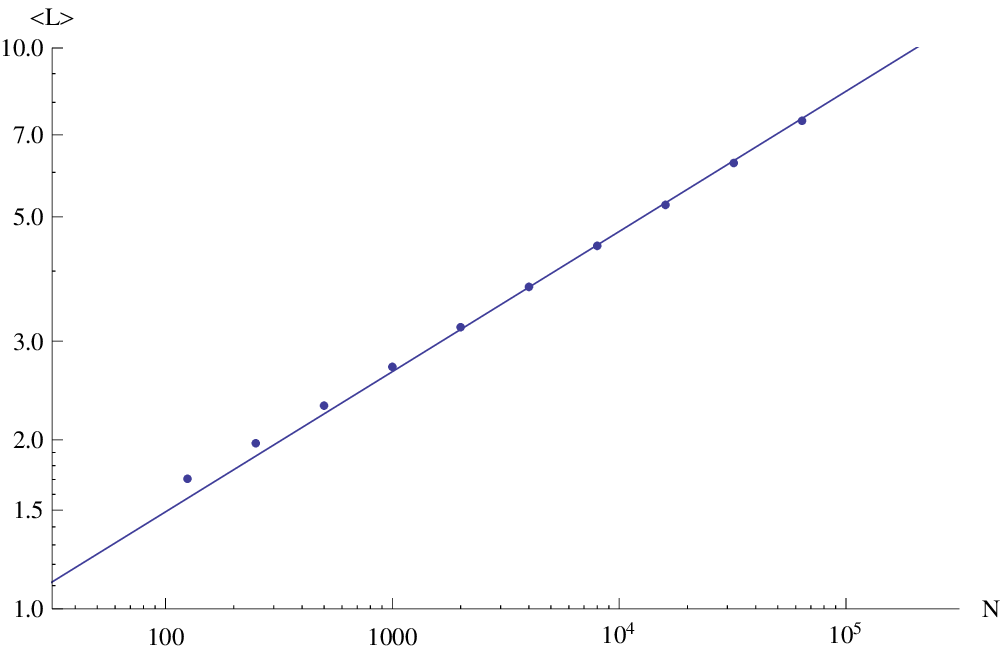}}}}
\caption[fig1]{Log-log-plot of the expectation value $\la L \ra_N$ versus $N$. 
The fitted curve is given by $\langle L \rangle_N = 0.471 N^{1/4}$. }
\label{dh4}
\end{figure}

Thus there seems to be a nice convergence to the correct 
distribution for $\tau_2$. However, let us try to understand 
in more detail the relation between the geometry of the triangulations
and the moduli parameter $\tau$. Due to (discrete) conformal invariance
the relation cannot be very direct. We observed above that the minimal 
length of a non-contractible loop was important for the 
$\tau$ distribution (for fixed $N$).
Let us classify a triangulation according to the length $L$ of 
its shortest non-contractible loop. In the piecewise linear geometry of $DT$ such
a loop will be a geodesic curve in the $DT$ sense. Thus we expect that 
its length $L$ scales anomalously with respect to the area $N$ 
\cite{fractal,fractal1}.
We can check this explicitly by determining shortest closed loops for random 
triangulations in the $DT$-ensemble using the method described in appendix \ref{shortestloop} (see also \cite{abbl}).
The result for the expectation value $\la L \ra_N$ is shown 
in Fig.\ \ref{dh4} and indeed we find that
\beq\label{1a}
\la L \ra_N \sim N^{1/d_h},~~~~d_h=4,
\eeq
where $d_h$ is the Hausdorff dimension of the $DT$-ensemble.
One can view the triangulation
to the left in Fig.\ \ref{reconstructionc-2} as 
an illustration of relation \rf{1a},
the length of the shortest non-contractible loop shown in red is 
of the order $N^{1/d_h}$, not of the order $\sqrt{N}$. Further 
implications of eq.\ \rf{1a} have been discussed in a recent paper \cite{abbl}.

\begin{figure}[t]
\centerline{\scalebox{1.0}{\rotatebox{0}{\includegraphics{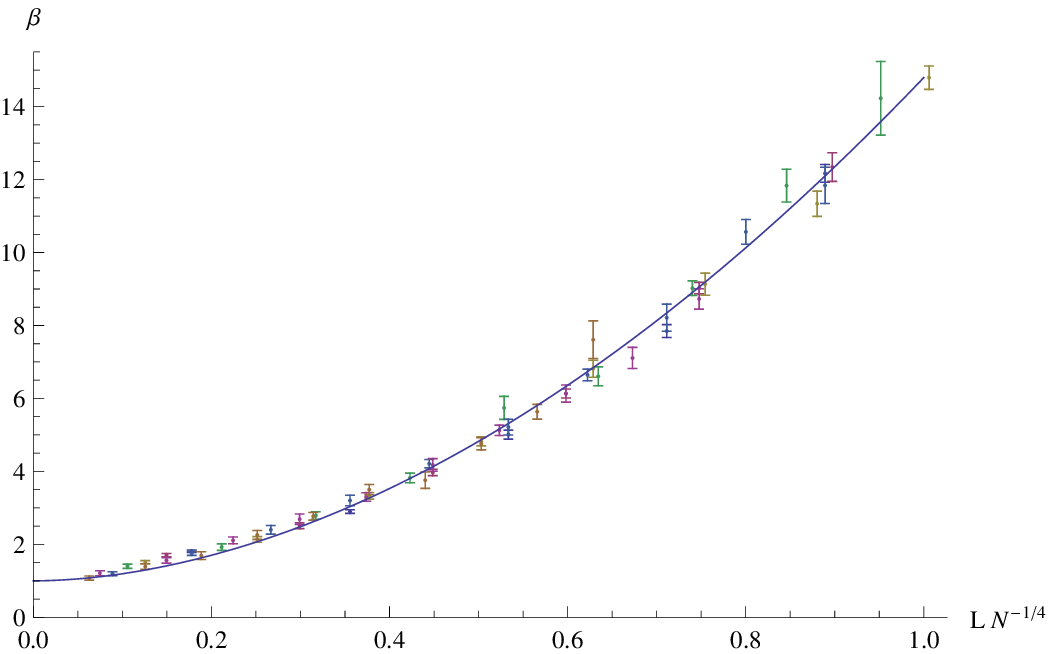}}}}
\caption{The exponential fall-off 
$P(N,L,\tau_2)\propto e^{-\pi \beta \tau_2/6}$ for $c=0$, 
formula \rf{2a}. The fit corresponds to 
$\beta = 1 + 13.7 \left(\frac{L}{N^{1/d_h}}\right)^{1.85}$.}
\label{beta}
\end{figure}

From the continuum formulas \rf{3s} and \rf{4s} we expect the $\tau$
distribution to fall off like $e^{-\pi \,\tau_2/6}$ for $\tau_2$ much larger
than 1. This is indeed in agreement with Fig.\ \ref{tau2-distribution}. 
However, if we split the $DT$-ensemble in subsets according to $L$, we
observe a {\it universal} dependence in terms of the ``dimensionless'' variable
$L/N^{1/d_h}$:
\beq\label{2a}
P(N,L,\tau_2) \sim \e^{ - \frac{\pi}{6} \,\tau_2\, \b},~~~~ 
\b= 1+13.7 \left(\frac{L}{N^{1/4}}\right)^{1.85} .
\eeq
The constants 13.7 and 1.85 are the result of a best fit
to the data (see Fig.\ \ref{beta}) and are not that precisely 
determined.
The important point (apart from the universality) is that the 
main contribution for large $\tau_2$ comes from small $L/N^{1/4}$.
Thus {\it statistically} there is a clear relation between small $L$'s
in the triangulation and large $\tau_2$'s in the embedded Euclidean 
plane. And ``small'' can be quantified as follows: since $d_h=4$ 
a typical linear extension of the triangulation, measured in geodesic 
distances, will be $N^{1/4}$, and ``small'' means small compared 
to this linear extension.
Finally, let us emphasize that the existence of a small $L$ 
in a triangulation does not {\it imply} that $\tau_2$ is large. 
For instance both independent holonomy directions could have
small non-contractible loops. As an extreme there could even 
be non-contractible loops of length one in both directions. 
In this case the embedding maps  (like the ones 
shown in Fig.\ \ref{reconstruction} and Fig.\ \ref{reconstructionc-2}) 
to the Euclidean plane could be 
extreme, having only a few large triangles and the rest
of the triangulation concentrated on a very little area. In such 
a case $\tau_2$ would not be large, rather we would just have 
a nice illustration of the power of local conformal transformations.
When $N$ is not too large we indeed observe such situations
since the smallest values of $L$ are not that rare.

\section{Measuring $\boldsymbol{\tau}$ for $\boldsymbol{c=-2}$}\label{c-2a}

We want to test eq.\ \rf{3s} for $c= -2$. As mentioned
$c=-2$ is chosen because it is easier to address numerically
than other matter systems coupled to two-dimensional Euclidean 
quantum gravity. The $c=-2$ matter coupled to quantum gravity has
several realizations. A minimal $(p,q)$-conformal field theory,
$p< q$, $p,q$ co-prime integers larger than 1, 
has central charge $c=1-6(p-q)^2/pq$ 
and, coupled to quantum gravity, the string susceptibility 
for surfaces with spherical topology is $\g_0(c)= 1-q/p$. Formally
a  $(p,q)=(1,2)$ model will have $c=-2$ and $\g_0=-1$. It
does not really belong to the minimal models and there is
no Kac table associated with $(p,q)=(1,2)$.
Nevertheless one can find a fermionic conformal field theory 
which formally can be identified with a (1,2) model \cite{distler}. 
It can (partly) be 
viewed as a topological field theory. This is the reason the 
$c=-2$ matter coupled to quantum gravity is often 
called topological 2d gravity. However, there exists another formal
representation of the $c=-2$ matter system coupled to 2d Euclidean 
quantum gravity. It also leads to $\g_0 = -1$ after integrating over
the matter fields, and in fact the value $\g_0=-1$ was first 
calculated using this representation in the $DT$ formalism \cite{kkm}.

Consider the partition function \rf{2.0} for the bosonic string
in $d$ dimensions. Since the $X_\m$ correspond to $d$ Gaussian fields
we can perform the integration over these fields and we will obtain
\beq\label{1b}
\int \cD_{\hg} X_\m \; e^{-S(\hg,X)} \sim \Big(\det(-\Del_\hg')\Big)^{-d/2}.
\eeq 
Here $\Del'_\hg$ denotes the Laplace-Beltrami operator on the background
geometry $\hg_{ab}$ and the prime signifies that the zero mode 
has been removed when calculating the determinant of $\Del_\hg$. If we 
consider $d$ as a formal parameter, $d=-2$ is special since
we just get the determinant itself, as one would have for a suitable
fermionic system, and since $d$ is the central charge for positive 
integer, we formally have a system where $c=-2$. The $DT$-formalism
tells us that we should represent the regularized partition function
\rf{2.0} as
\beq\label{2b}
Z^{(h)}(\m) = \sum_T  \frac{1}{C_T}\;\e^{-\m N(T)} {\det}_T (-\Del_T'),
\eeq
where the summation is over all triangulations of genus $h$ in a suitable
$DT$-ensemble. $\m$ denotes the bare cosmological constant of the 2d 
quantum gravity theory (the Liouville cosmological constant in the 
continuum notation), $N(T)$ the number of triangles in $T$ and 
$C_T$ is symmetry factor of the triangulation $T$, i.e.\ 
the order of the automorphism group of the graph $T$.
Finally $\Del_T$ denotes the (discretely defined) Laplacian on 
the $DT$-surface, which we take to be the usual graph Laplacian of 
the $\phi^3$ graph dual to the triangulation. 
We can trade the cosmological constant $\m$ for the area of the 
(triangulated) surface. Since each triangle has the same area 
in the $DT$ formalism, this implies keeping the number of triangles 
fixed and our partition function becomes
\beq\label{3b}
Z^{(h)}(N) = \sum_{T \in DT(N)} \frac{1}{C_T}\; {\det}_T  (-\Del_T'),
\eeq
where $DT(N)$ denotes the $DT$-ensemble with $N$ triangles. 
We expect $Z^{(h)}(N)$ to approach the continuum $Z^{(h)} (A)$
for $N \to \infty$ via the identification $A = N a^2 \sqrt{3}/4$,
where $a$ denotes the length of the links in the triangulation.

It is well-known that the determinant of a graph Laplacian 
(with the zero-mode removed) 
is equal to the number of spanning trees on the graph.\footnote{For a given 
connected graph, a spanning tree is a connected subgraph
which contains all the vertices, but which contains 
no loops, i.e. closed paths of links.} Therefore $\det_T(-\Del_T')=\cN(T)$ 
where $\cN(T)$ denotes the number spanning trees of the $\phi^3$ graph dual 
to the triangulation $T$. It was this representation
of the combinatorial Laplacian which allowed to authors of \cite{kkm}
to solve the model for $h=0$ and prove that $\g_0=-1$, a result 
which was controversial at the time it was published. 

The partition function becomes
\beq\label{4b}
Z^{(h)} (N) = \sum_{T\in DT(N)} \frac{1}{C_T}\;\cN(T) = \sum_{T_{N}(ST)} 
\frac{1}{C_T},
\eeq
where $T_{N}(ST)$ denotes 
the set of genus $h$ triangulations which are ``decorated''
by spanning trees. Thus two such triangulations are
counted as different even if they as triangulations are identical,
but if they have different ``embedded'' spanning trees.  
 
In \cite{kk} it was realized that this formula, in the case of genus
zero ($h=0$), could be used in computer simulations to generate 
directly, by a recursive algorithm, a set of graphs $T$ which 
have the correct weight, including the matter fields. Thus, in 
order to test the properties of $c=-2$ matter coupled to 2d 
quantum gravity, one can directly generate a set of independent
graphs of arbitrary size and circumvent the problem of generating
statistically independent configurations by local Monte Carlo updating,
a problem which becomes increasingly time consuming when $N$ is 
large. This way of dealing numerically with the $c=-2$ system
has been used extensively to study the fractal properties of 
quantum gravity \cite{c-2fractal}. However, both the original solution
of the model and the successive applications used in a crucial 
way that $h=0$. For our application we need a generalization
of the $h=0$ method to higher genera, or at least to genus one.
Such a method is described for the simplest case $h=1$ in the next 
section using a very recent result from \cite{chapuy2011}. 
In appendix \ref{highergenus} we sketch the procedure for larger genus.

\subsection{Random generation of decorated torus triangulations}\label{c-2method}
 
We will now describe an algorithm which generates a random triangulation 
$T$ of genus $h$ with $N$ triangles decorated with a spanning tree. 
The algorithm is designed such that any decorated triangulation is 
sampled with equal probability. The class of triangulations we are 
considering is the most general one, i.e. we allow for triangles to 
be glued to themselves and two different triangles are allowed to be 
glued along more than one edge as long as the triangulation remains 
connected and the resulting piecewise linear manifold remains homeomorphic
to a surface of genus $h$.  

\begin{figure}[t]
\centerline{\scalebox{0.7}{\rotatebox{0}{\includegraphics{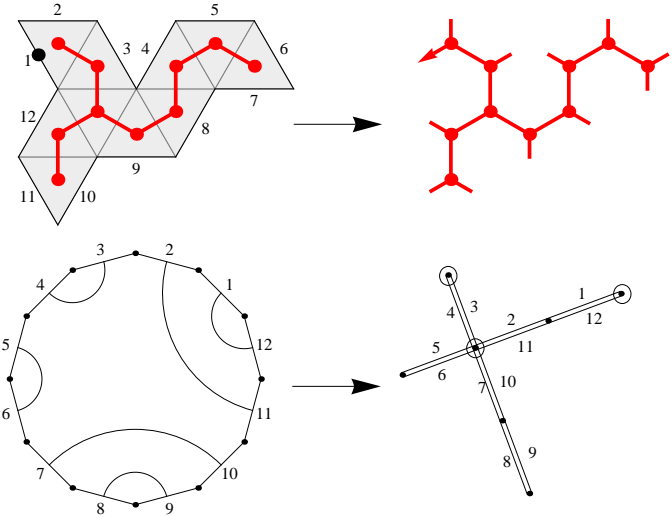}}}}
\caption{The ingredients that specify a decorated genus 0 triangulation 
with $N=10$ triangles. The top figure shows how a triangulated disc 
with a marked boundary edge corresponds to a rooted binary tree. 
The bottom figure shows how the pair-wise gluing 
$\alpha = \{\{1,12\},\{2,11\},\{3,4\},\{5,6\},\{7,10\},\{8,9\}\}$ 
of the edges corresponds to a unicellular map which for genus 0 is 
just a planar tree with a marked half-edge (i.e. the one labeled 1). 
To construct a genus 1 unicellular map we select three vertices in 
the planar tree (the encircled ones). The distinguished half-edges 
are $a_1=2$, $a_2=3$ and $a_3=12$. Therefore we should relabel 
$\{1,2,3,4,5,6,7,8,9,10,11,12\}\to\{1,2,12,3,4,5,6,7,8,9,10,11\}$ 
in $\alpha$, yielding the new pair-wise gluing
$\{\{1,11\},\{2,10\},\{3,12\},\{4,5\},\{6,9\},\{7,8\}\}$.\label{fig:map}}
\end{figure}

Given a triangulation together with a spanning tree on its dual graph,
we consider the set of edges of the triangulation 
which are not intersected by the spanning tree.
This set of edges forms a graph consisting of $N/2 +1$ 
links and containing $2h$ loops. 
If we cut open the triangulation along these links, we obtain
a triangulation of the disc with a boundary consisting of $N+2$ links. 
This triangulated disc is completely characterized by the structure 
of a trivalent tree with $N$ internal vertices 
(see the top part of Fig. \ref{fig:map}). 
To get back to the original triangulation, the boundary 
edges of the disc have to be glued pairwise. 
This suggests that any decorated triangulation can be 
obtained by combining a tree and a pairwise gluing.

Let us make this a bit more precise. We can view the
$N$ vertices of the spanning tree as located at the center of the triangles. 
The order of the vertices (the number of links to which they belong)
can be one, two or three. By construction they are located at 
triangles where two, one or no links have been cut open. Let us 
add two, one or no ``external'' links to the vertices. 
One can visualize this as the external lines still being located
in the triangle to which the spanning tree vertex belongs, and 
``pointing'' to the triangle links which have been cut open. In this 
way the spanning tree has been extended to a tree where the  
$N$ vertices have become internal vertices of order three 
and where $N+2$ external links have been added.
We mark one of the external links in order to have a rooted tree. 
In this way the tree becomes precisely of the form of a {\it binary tree}.
The external links are now, by construction, in 1-to-1 correspondence
with the boundary links of the disc. The pairwise gluing of the 
edges of the disc to a genus $h$ surface corresponds to 
a pairwise identification of the external lines in the binary 
tree such that the resulting trivalent graph can be placed
on a genus $h$ surface without any lines crossing (this is 
the meaning of the trivalent graph being of genus $h$).
Such a pairwise gluing of a polygon is known as a unicellular map \cite{chapuy2011} (or one-face map) 
of genus $h$ with $N+2$ ``half-edges''.

The above arguments show that we can generate a random decorated triangulation 
by separately generating a random trivalent planar tree and a random 
unicellular map. We then use the unicellular map to connect the 
external lines of the trivalent tree. 

As already remarked the random trivalent planar trees 
with a marked external line are in 1-to-1 correspondence with binary trees.
There exists efficient algorithms to generate such trees, 
see e.g. \cite{knuth2005} section 7.2.1.6.

The major problem is to implement the random unicellar map. 
The $c=-2$ model was originally solved for $h=0$ because
it was understood that a genus zero unicellular map 
with $N+2$ half-edges is given simply by a planar tree with 
one half-edge marked \cite{kkm}. Fig.\ \ref{fig:map} illustrates this: 
the identification 
of the half-edges has to form ``rainbow'' diagrams in order
that one creates a genus zero surface. The corresponding planar 
trees are again related to binary trees and can thus be 
easily generated randomly.

To generate a torus we need a random genus one unicellular map 
with $N+2$ half-edges. Luckily in \cite{chapuy2011} an explicit 
connection was found between unicellular maps of genus $g$ and 
genus $g+1$. In particular, for a genus zero unicellular map a procedure 
is given in which three distinct vertices are identified and the 
half-edges are relabeled in such a way that one obtains a genus one 
unicellular map. It is shown that any genus one unicellular map can be 
obtained through such a procedure in exactly two different ways 
(see \cite{chapuy2011}, proposition 1 and corollary 1).

Let us briefly summarize the procedure (see Fig. \ref{fig:map} for 
an example). We label the half-edges of the $N+2$-gon anti-clockwise 
by $1,2,\ldots,N+2$ and we provide them with the corresponding orientation, such that they have
a starting vertex and a final vertex. 
A unicellular map of genus zero is fixed by 
giving a list $\alpha$ of $N/2+1$ pairs of integers which tell 
us which edges to glue. After the gluing 
we have a tree with $N/2+2$ vertices (see Fig.\ \ref{fig:map}) 
of which we randomly select three distinct ones. For each of them we 
select from the set of half-edges having that vertex as its {\it final}
vertex  the smallest index. 
We denote these indices by $a_1$, $a_2$, and $a_3$ and reorder them 
such that $1 \leq a_1 < a_2 < a_3 \leq N+2$. The resulting unicellular 
map of genus one is now given by gluing according to $\alpha$ in which 
we replace $i \to f(i)$, where 
\begin{equation}\label{5b}
f(i) = \left\{ \begin{array}{ll} i + a_3 - a_2 & 
\mathrm{if }\, a_1<i\leq a_2\\ i - a_2 + a_1 & \mathrm{if }\,a_2<i\leq a_3 \\ 
i & \mathrm{otherwise}\end{array}\right.
\end{equation}
We refer the \cite{chapuy2011} for the actual proof of this statement.

\subsection{Numerical results for $\boldsymbol{c=-2}$}\label{c-2b}

\begin{figure}[t]
\centerline{\scalebox{1.0}{\rotatebox{0}{\includegraphics{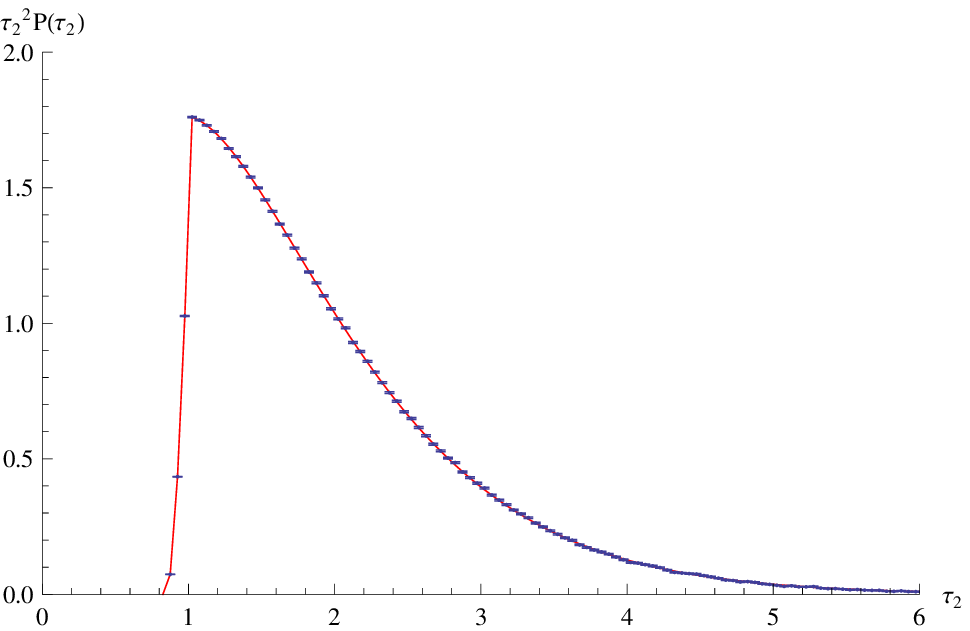}}}}
\centerline{\scalebox{1.0}{\rotatebox{0}{\includegraphics{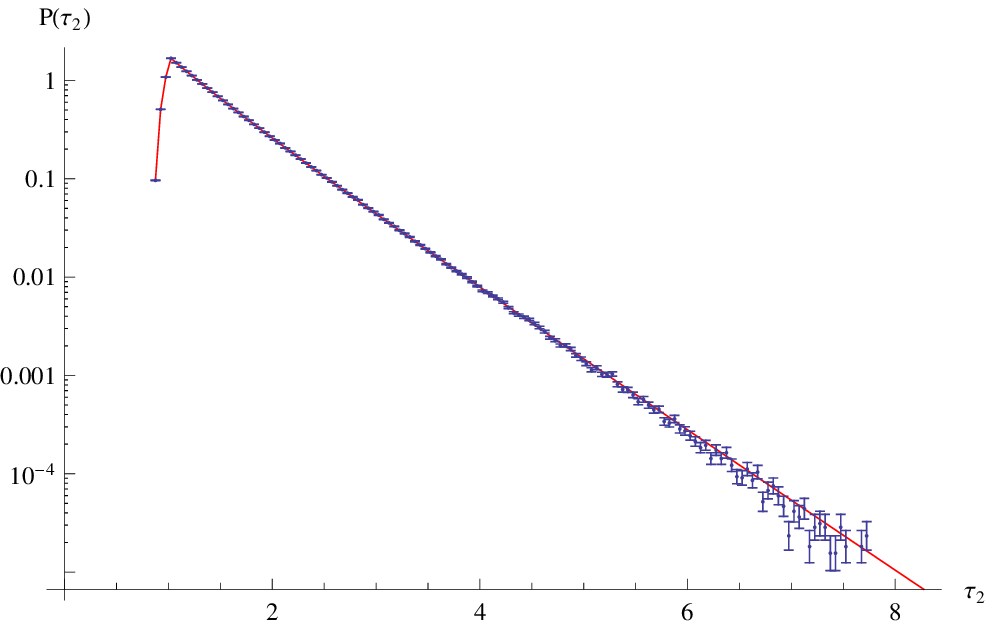}}}}
\caption{The $\tau_2$ distribution for $N=8000$,
compared to the theoretical distribution (the red curve). }
\label{tau2c-2}
\end{figure}

We use the above described algorithm to generate an ensemble of 
$c=-2$ graphs and we use this to perform the same measurements 
as for $c=0$. Let us record the $\tau_2$ distribution. It is 
shown in Fig.\ \ref{tau2c-2}.
We have a perfect fit even for a relatively small 
triangulation of 8000 triangles. In fact we have seen no deviation from the 
theoretical curve not compatible with the error-bars, so we have 
not addressed the approach to the theoretical curve as a function 
of $N$ as we did for $c=0$. 

\begin{figure}[t]
\centerline{\scalebox{1.0}{\rotatebox{0}{\includegraphics{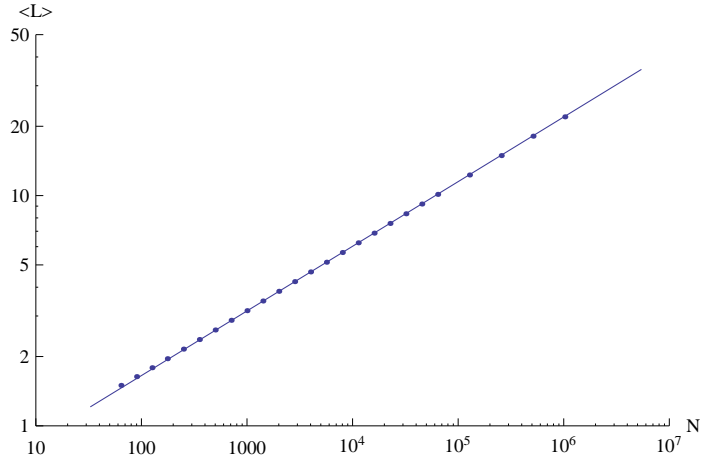}}}}
\caption{Log-log-plot of the expectation value $\langle L\rangle_N$ of the length of the shortest closed loop as a function of the volume $N$. The fitted curve corresponds to $\la L \ra_{N} = 0.454 N^{1/3.56}$ (error-bars too small to display).} 
\label{dh-2}    
\end{figure}

\begin{figure}[ht]
\centerline{\scalebox{1.0}{\rotatebox{0}{\includegraphics{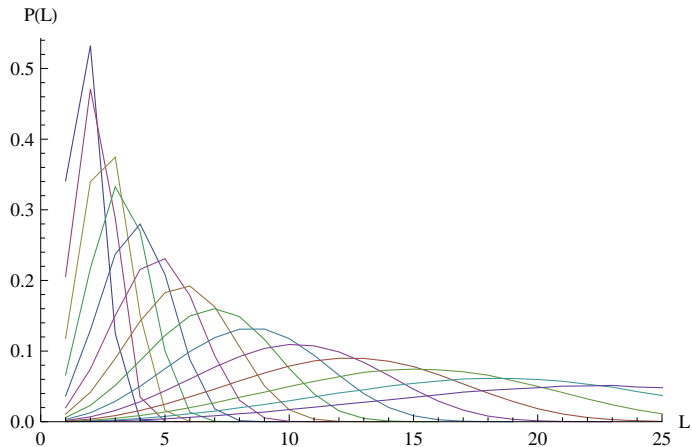}}}}
\caption{The distributions $P_N(L)$ for $N=126$ up to 1024000 (the error bars are too small to display).}
\label{pnl} 
\end{figure}

\begin{figure}
\centerline{\scalebox{1.0}{\rotatebox{0}{\includegraphics{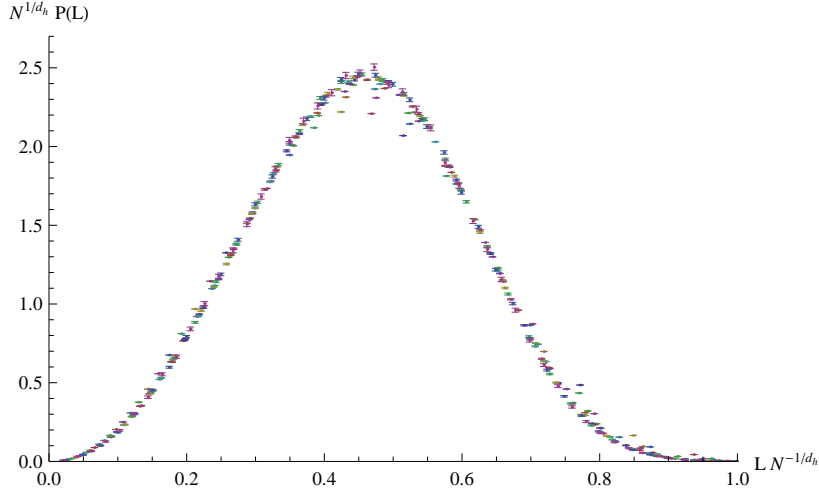}}}}
\caption{The rescaled distribution
$\tilde{P}(x)$ defined in \rf{7b} for $N=126$ up to 1024000.}
\label{fssc-2}
\end{figure}

We measured the expectation value $\langle L \rangle_N$ of the length of the shortest non-contractible 
loop and the result is shown in Fig.\ \ref{dh-2}.
As for $c=0$ we expect anomalous scaling according to \rf{1a}, only now $d_h$ is no longer equal to 4, but given by the formula \cite{watabiki}
\beq\label{6b}  
d_h(c) = 2 \frac{\sqrt{49-c}+\sqrt{25-c}}{\sqrt{1-c}+\sqrt{25-c}},
\eeq
for $c\leq 1$. The data shown in Fig.\ \ref{dh-2} is in 
perfect agreement with \rf{6b}, which for $c=-2$ 
becomes $d_h = (3+\sqrt{17})/2\approx 3.56$. 
Eq.\ \rf{6b} was earlier 
verified with good precision for genus zero surfaces by directly 
measuring the area $N(r)$ enclosed within a circle of geodesic radius $R$
and showing that $\la N(R) \ra_R \sim R^{d_h}$, with $d_h$ given by 
\rf{6b}~\cite{c-2fractal}.
As already noticed, \rf{1a} can be viewed as an independent verification
of the anomalous scaling of geodesic distance. Another rather 
stunning confirmation that the intrinsic structure of 2d 
quantum gravity is governed by the dimensionless quantity 
$R/A^{1/d_h}$, $A$ being the area of the 2d universe, $R$ a geodesic
distance, can be obtained
by looking closer at the actual probability distribution $P_{N}(L)$ 
of the length of the shortest non-contractible loop as a function of $N$. 
In Fig.\ \ref{pnl} we have shown the distribution for
a range of $N$ stretching from 126 to more than $10^6$. 
In Fig.\ \ref{fssc-2} we  
show that all of these $P_{N}(L)$ are well described by 
\beq\label{7b}
P_{N}(L) = N^{1/d_h} \, \tilde{P}(x),~~~~x=\frac{L}{N^{1/d_h}}.
\eeq 
This finite size scaling over such an amazing range of $N$'s explains
why we could hardly see any deviation from the continuum theoretical 
result in Fig. \ref{tau2c-2}: already $N=8000$ is in some sense 
very close to the $N= \infty$ limit according to \rf{7b}. 

\begin{figure}[ht]
\centerline{\scalebox{1.0}{\rotatebox{0}{\includegraphics{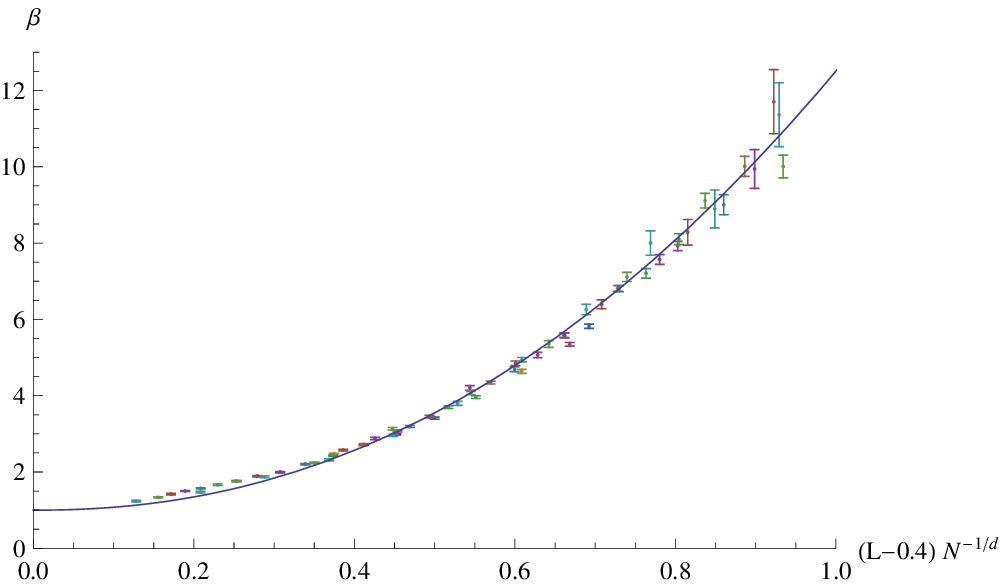}}}}
\caption{The exponential fall-off 
$P(N,L,\tau_2)\propto e^{-\pi \beta \tau_2/2}$ for $c=-2$, formula \rf{8b}. 
The fit corresponds to 
$\beta = 1 + 11.4 \left(\frac{L-0.4}{N^{1/d_h}}\right)^{2.2}$.}
\label{betac-2}
\end{figure}

For a conformal field theory with central charge $c$ 
we expect from the continuum formulas \rf{3s} and \rf{4s} that the  $\tau$
distribution will fall off like $e^{-\pi(1-c) \,\tau_2/6}$ for $\tau_2$ much larger
than 1. We already checked this for $c=0$ and
from Fig.\ \ref{tau2c-2} it is clear that it is also true for $c=-2$.
Again, if we split the $DT$-ensemble in subsets according to $L$, 
we observe a {\it universal} dependence in terms of the ``dimensionless'' 
variable $L/N^{1/d_h}$ (Fig.\ \ref{betac-2}):
\beq\label{8b}
P(N,L,\tau_2) \sim \e^{ - \frac{\pi}{2} \,\tau_2\, \b},~~~~ 
\b= 1+11.4 \left(\frac{L-\a}{N^{1/4}}\right)^{2.2} .
\eeq
The constants $11.4$ and $2.2$ are different from the $c=0$ ones,
and we have included a ``shift''\footnote{The ``shift'' $\a$ can be 
viewed as a simple
way to compensate for discretization effects for small $L$'s. We expects 
a formula like \rf{2a} or \rf{8b} 
to reflect a continuum dependence $\ell/A^{1/d_h}$
where $\ell$ is the continuum length $\ell = L a$ and $A$ is 
the continuum area $A \propto N a^2$. While $\ell$ can 
be arbitrarily small, this is of course not the case for $L$
which is an integer. A priori there will be discretization 
effects if $L$ is not much larger than 1. However, it is know that 
a shift $L/N^{1/d_h} \to (L-\a)/N^{1/d_h}$ can reduce the 
discretization effects \cite{fractal1,fractal2} and for $c=-2$ it 
does improve the fit.  For $c=0$ the shift is not important and
we left it out in \rf{2a}.} in the integer values $L$ by
$\a=0.4$, but the 
message is the same: at the boundary $\tau_2=\infty$ 
of moduli space the $\tau$ distribution is completely determined by
the smallest $L$'s, where the $N$ independent statement of ``small'' 
is that $L/N^{1/d_h}$ is small.   

\section{Discussion}

Starting from a path integral approach to two-dimensional 
Euclidean quantum gravity coupled to matter fields, we need
a regularization of the path integral if we want to have 
a definition of the theory which is not only formal. The formalism 
of dynamical triangulation is such a regularization. One can
view the $DT$-formalism in two ways. In the first 
a  triangulation represents a continuum 
piecewise linear geometry. Thus the geometry is
considered  flat except at the vertices. At the vertices geometric
quantities such as curvature, which involve derivatives of the
geometry, can be singular. In a natural way the curvature
has a delta-function distribution, being located at the vertices. 
One can take the point of view that the vertex represents a conical 
singularity with a certain deficit angle, but it might not be important
to take such a literal continuum interpretation. At least it is a cumbersome
road to take since we are interested in a limit where the number 
of vertices goes to infinity, and since we consider the triangles as
{\it building blocks}, keeping them all as equilateral triangles, 
the deficit angles are not going to zero even if we consider triangulations
with an increasing number of building blocks and even if we rescale 
all side lengths $a$ of the triangles to zero. 
Alternatively we might simply consider a triangulation as a lattice 
realization of a 2d geometry, where not too much emphasis should be 
put on the piecewise linear structure.

The analogy with the 
textbook derivation of the ordinary path integral in quantum mechanics 
might be useful: the path integral $\int \cD x(t)\cdots $ is obtained
by discretizing the time interval, which leads to a multidimensional 
integral $\int\prod_{i=1}^N dx(t_i)\cdots$. We can now {\it choose} to 
view the points $x(t_i)$ as vertices in a piecewise linear path $x(t)$
connecting the initial point $x_i$ and the final point $x_f$. With
such a choice, the piecewise linear paths with $N$ vertices becomes a subset of the full set of continuous paths
connecting $x_i$ and $x_f$,  which enter into 
the continuum path integral. It can be shown that this 
subset of piecewise linear paths is a dense set in the limit $N\to \infty$
when the right norm is used to define distances between continuum paths,
namely the norm compatible with the Wiener measure (see \cite{book} for 
a detailed discussion). However, we are not forced to take such a 
point of view for the particle path integral. We could simply 
view $\prod_{i=1}^N dx(t_i)\cdots$ as  a lattice version of
the formal continuum expression and whatever action 
we use in the continuum, we choose a suitable discretized lattice 
version of derivatives etc. Then, when the lattice spacing goes
to zero we expect to obtain the continuum theory. Universality
in the Wilsonian sense is then the key ingredient for obtaining
a universal continuum limit. For the free relativistic particle 
this was analyzed in detail in \cite{book} and the universality can be 
shown to be a consequence of the central limit theorem.

Presently we have no rigorous mathematical 
definition of the formal measure $\int \cD [g]$ over {\it geometries}
but it is natural to believe that it involves the integration over 
all continuous geometries\footnote{Clearly, using the 
standard, formal, continuum procedure summarized in \rf{2.0},
one ends up with an integration over the 
field $\phi$.  The path integral over $\phi$ clearly involves 
fields $\phi$ which are continuous but nowhere 
differentiable. The same will then be true for the 
corresponding geometries defined by the metric 
$g_{\a\b} = \e^{\phi}\hg_{\a\b}$, where $\hg_{\a\b}$ is a 
fixed background metric.}. If we take the first point of view 
advocated for the path integral of the particle, we can choose 
to consider the $DT$-ensemble 
as a subset of the continuous geometries which, when the number 
of triangles $N$ goes to infinity  while the length $a$ of the links goes to 
zero in such a way that the 
$V= Na^2$ is kept fixed, hopefully becomes dense in the set of 
continuous geometries with volume $V$. However, contrary
to the situation for the particle path integral, we cannot 
{\it prove} this since   
we do not presently know the norm which defines distances 
in the space of continuous two-dimensional geometries.
As mentioned in the Introduction there is indeed some
evidence that the $DT$-set of piecewise linear geometries 
can be viewed as dense in the set of continuous geometries,  
since a number of global quantities calculated using the 
formal continuum path integral agree with the corresponding quantities
calculated using the $DT$-ensemble.

However, the actual use of the $DT$-ensemble, both in analytical calculations
and numerical simulations, is more in the spirit 
of the second point of view advocated for the path integral of the 
particle. According to this point of view we consider
the $DT$-formalism as some conveniently chosen 
UV  lattice representation of the 
two-dimensional geometries,
where one should not put too much emphasis on the detailed piecewise
linear geometric interpretation. Indeed, in the actual numerical
simulations one measures geodesic distance
using lattice links or dual links, not the actual geodesic 
distance in the piecewise linear geometry. Also, one is using discrete
differences as substitutes for derivatives, rather than trying 
to formulate the field theory on a geometry with conical singularities.
Similarly, one often considers discrete spin systems, like the Ising 
model, representing  some  conformal field theories. Such systems
are of course {\it only} defined on a lattice. All numerical evidence 
as well as analytical calculations support the Wilsonian idea that 
lattice details are relatively unimportant when taking the continuum limit. 

It is with this second point of view 
in mind that we have analyzed the moduli structure
provided to us by the $DT$-ensemble. We have not tried to view the 
triangulation as piecewise linear geometry with conical singularities
to which one can associate a precise conformal structure
as for instance described in \cite{carfora}. Rather,
we have just imitated a standard analytic construction of the moduli
parameter $\tau$ by replacing the harmonic  differentials with 
suitable discrete differentials. This is in the spirit of the 
lattice approach. The conclusion: it works almost perfectly. 
In particular the $c=-2$ results, where we were not depending 
on Markov chain Monte Carlo simulations, are in perfect agreement with the 
continuum results. Our numerical experiments also provide us 
with additional evidence that the anomalously scaling geodesic 
distance plays a key role in any deeper understanding of 
observables in quantum gravity theories.


\subsection*{Acknowledgments}
 JA would like to thank the Institute of 
Theoretical Physics and
the Department of Physics and Astronomy at Utrecht University for hospitality 
and financial support. He also acknowledge financial 
support by  the Danish Research Council (FNU) from the grant 
``quantum gravity and the role of black holes''.
TB acknowledge support by the Netherlands Organisation for Scientific 
Research (NWO) under their VICI program.

\appendix
\section{Generalization to higher genus}\label{highergenus}

The way we have set up the moduli measurements in section \ref{setup} 
allows for a rather straightforward generalization to surfaces of genus 
$h$ larger than 1.\footnote{A method similar to the one described here 
has been previously employed in a completely different context in \cite{gu}.} 
For such surfaces the complex structure is characterized by the 
Teichm\"uller space of complex dimension $3 h-3$. 
Identifying explicitly $3 h-3$ complex moduli parameters is a hard 
task and one which we will not pursue for general genus. Instead we 
will use a direct generalization of the genus 1 moduli parameter 
$\tau$ in the complex upper half-plane $\mathbb{H}$ to matrices 
$\Omega$ in the $h$-dimensional Siegel upper half-plane $\mathbb{H}_h$, known as period matrices \cite{farkas}. The Siegel complex upper half-plane 
$\mathbb{H}_h$ consists of all symmetric complex $h$ by $h$ matrices 
$\Omega$ that have a positive-definite imaginary part. A period matrix 
corresponding to a surface is known to completely determine its complex 
structure. For genus $h\leq 2$ we can actually identify $\mathbb{H}_h$ 
with Teichm\"uller space, but for $h\geq 3$ Teichm\"uller space appears as a 
non-trivial submanifold sitting in $\mathbb{H}_h$ (as is apparent from 
comparing their dimensions).

In order to define the period matrix for a Riemannian manifold we have to 
first choose a set of $2h$ closed curves $a_1,\ldots,a_h,b_1,\ldots,b_h$ 
which generate the fundamental group and which satisfy
\begin{equation}\label{intersection}
i(a_i,a_j)=i(b_i,b_j)=0 \quad \mbox{and} \quad i(a_i,b_j)=\delta_{ij}
\end{equation}
where $i(\cdot,\cdot)$ denotes the oriented intersection number. The 
space of harmonic 1-forms is  $2h$-dimensional and it is possible to 
choose a basis $\{\alpha_i,\beta_i\}$ dual to the curves $a_i$ and 
$b_i$ in the sense that
\begin{equation}
\int_{a_i}\beta_j=\int_{b_i}\alpha_j=0 \quad \mbox{and} 
\quad \int_{a_i}\alpha_j =\int_{b_i}\beta_j=\delta_{ij}.
\end{equation}
A complex basis of holomorphic 1-forms is given by
\begin{equation} 
\omega_i = \alpha_i + i\; {\ast \alpha_i},
\end{equation}
where $\ast$ is the Hodge dual. The period matrix $\Omega$ is then 
given by $\Omega = A^{-1}B$ in terms of the matrices
\begin{equation}
A_{ij} = \int_{a_j}\omega_i  \quad \mbox{and} \quad B_{ij} = \int_{b_j}\omega_i.
\end{equation}
We can express these integrals in terms of the inner-products 
$\langle \alpha_i|\alpha_j\rangle$, $\langle \alpha_i|\beta_j\rangle$ 
and $\langle \beta_i|\beta_j\rangle$ by using the Riemann bilinear 
relations \cite{farkas}, which state that any two closed 1-forms 
$\rho$ and $\sigma$ satisfy
\begin{equation}\label{rbrel}
\int \rho \wedge \sigma = \sum_{i=1}^h \int_{a_i} \rho \int_{b_i} 
\sigma - \int_{a_i} \sigma \int_{b_i} \rho.
\end{equation}
Therefore
\begin{equation}
A_{ij} = \delta_{ij} + i \int_{a_j}\ast\alpha_i = 
\delta_{ij} + i\int \ast \alpha_i \wedge \beta_j = 
\delta_{ij} - i \langle \alpha_i | \beta_j\rangle
\end{equation}
and 
\begin{equation}
B_{ij} = i \int_{b_j}\ast\alpha_i = 
i \int \alpha_j \wedge \ast \alpha_i = i\langle \alpha_i | \alpha_j\rangle.
\end{equation}
Another consequence of (\ref{rbrel}) is that
\begin{eqnarray}
\delta_{ij}&=&\int \alpha_i\wedge\beta_j = 
\int \ast\alpha_i\wedge\ast\beta_j = 
\sum_{k=1}^h\int_{a_k}\ast\alpha_i\int_{b_k}\ast\beta_j-
\int_{a_k}\ast\beta_j\int_{b_k}\ast\alpha_i \nonumber\\
&=& \sum_{k=1}^h\int\beta_k\wedge\ast\alpha_i\int\alpha_k\wedge\ast\beta_j-
\int\beta_k\wedge\ast\beta_j\int\alpha_k\wedge\ast\alpha_i\nonumber\\
&=&\sum_{k=1}^h \langle\alpha_i|\alpha_k\rangle\langle\beta_k
|\beta_j\rangle-\langle\alpha_i|\beta_k\rangle\langle\alpha_k|
\beta_j\rangle\label{normalization}.
\end{eqnarray}

These formulae we can directly apply to $DT$ piecewise linear geometries 
by replacing $\alpha_i$ and $\beta_i$ by their discrete counterparts 
and the inner-product by the discrete one from (\ref{10s}).
For genus $h=1$ we used (\ref{normalization}) to establish the 
normalization of the discrete inner product leading to the expression 
for $\Omega = [\tau]$ in (\ref{9s}).
For genus $h\geq 2$ there is an overall ambiguity in the definition of 
$\Omega$ because (\ref{normalization}) is not exactly satisfied 
(up to overall factor) due to discretization artefacts.
However, we expect (and we have confirmed this numerically for genus 2) 
that for large random surfaces (\ref{normalization}) is close to a 
multiple of the identity matrix with high probability.
In that case we can unambiguously normalize the inner-product and 
determine $\Omega$.

The modular group $SL(2,\mathbb{Z})/\mathbb{Z}_2$ which acts on the upper 
half-plane as in (\ref{30s}) generalizes to the action of the symplectic 
group $Sp(2h,\mathbb{Z})/\mathbb{Z}_2$ on the Siegel upper half-plane 
$\mathbb{H}_h$. Fundamental domains can again be worked out but become 
increasingly cumbersome for larger genus (see e.g. \cite{geer}).

Finally let us mention that our algorithm for generating random $DT$ 
surfaces coupled to $c=-2$ conformal matter can be straightforwardly 
extended to genus $h\geq 2$. The only missing ingredient is a 
construction of a random unicellular map of genus $h$. Here again we 
can use results from \cite{chapuy2011} in which for any genus $h$ an 
explicit bijection is found between the set of unicellular maps of 
genus $h$ and a union of sets of unicellular maps of lower genus with a 
particular number of distinguished vertices.

\section{Finding shortest non-contractible loops}\label{shortestloop}

First let us present a method of constructing curves in the triangulation that generate the fundamental group. Inspired by the methods described in Sec. \ref{c-2method} we generate an arbitrary spanning tree on the $\phi^3$ graph dual to the triangulation. Then we consider the set of edges of the triangulation that are not intersected by this spanning tree.\footnote{Notice that in the case of randomly generated triangulations coupled to $c=-2$ conformal matter we have these ingredients already by construction.} As mentioned in Sec. \ref{c-2method} the graph $G$ formed by these edges contains $2h$ cycles, where $h$ is the genus of the triangulation. To extract these cycles we choose an arbitrary vertex $v$ in $G$ and generate a spanning tree for $G$ based at $v$. This tree will contain all but $2h$ of the edges of $G$. Adding any of the remaining $2h$ edges to the tree will lead to a cycle and therefore to a unique closed path based at $v$. By rearranging (and if necessary composing) the paths thus obtained we arrive at a canonical set of generators $\{ \gamma_i \}_{i=1,\ldots,2h}$, i.e. a set that satisfies (\ref{intersection}), for the fundamental group of the triangulation.

Once we have such a set of generators we can establish for any closed curve whether it is contractible or not by computing its oriented intersection number with the generators $\gamma_i$. The curve is contractible if and only if all these intersection numbers vanish. To make this test more efficient notice that we can easily construct by hand for each generator $\gamma_i$ a closed discrete 1-form $\phi_i$, i.e. satisfying $d\phi_i=0$, such that the intersection number of a curve with $\gamma_i$ is equal to the discrete integral of $\phi_i$ along that curve. Hence, a closed curve is non-contractible if the integral of at least one of the 1-forms $\phi_i$ is non-vanishing.

Given a vertex $v$ we can find a shortest non-contractible loop based at $v$ by performing a so-called breadth-first search in the edge-graph of the triangulation starting at $v$. Once we encounter a vertex that we have already visited before, we have implicitly established a loop in the edge-graph. The first such loop we meet that is non-contractible will automatically have minimal length. 

Now in principle we can repeat this procedure for each vertex $v$ in the triangulation to find the overall shortest non-contractible loop (or rather a non-contractible loop of minimal length as there usually more than one). However in general the set $V$ of vertices for which we have to perform this procedure can be greatly reduced. Indeed, we known that any non-contractible loop will intersect at least one of the generators $\gamma_i$, so it suffices to take $V$ to consist of all vertices contained in the $\gamma_i$. In order to obtain such a set $V$ with as few vertices as possible it is worthwhile to first spend some time to shorten the $\gamma_i$. This will result in a set $V$ with a number of vertices of the order $N^{1/d_h}$ with $N$ the number of triangles. Since a single breadth-first search involves a number of steps of the order $N$, the full algorithm will have an expected run-time of the order $N^{1+1/d_h}$, which amounts to $N^{1.25}$ for $c=0$ and $N^{1.281}$ for $c=-2$.


\begin{thebibliography}{99}

\bibitem{ackm}
  J.~Ambjorn, L.~Chekhov, C.~F.~Kristjansen, Y.~Makeenko,
  Nucl.\ Phys.\  {\bf B404 } (1993)  127-172.
  [hep-th/9302014].
\bibitem{ak}
  J.~Ambjorn, C.~F.~Kristjansen,
  Mod.\ Phys.\ Lett.\  {\bf A8 } (1993)  2875-2890.
  [hep-th/9307063].

\bibitem{bertrand}
  L.~Chekhov, B.~Eynard, N.~Orantin,
  JHEP {\bf 0612 } (2006)  053.
  [math-ph/0603003].\\
  L.~Chekhov, B.~Eynard,
  JHEP {\bf 0603 } (2006)  014.
  [hep-th/0504116].


\bibitem{cylinder}
  E.~J.~Martinec,
  [hep-th/0305148].\\
  N.~Seiberg, D.~Shih,
  JHEP {\bf 0402 } (2004)  021.
  [hep-th/0312170].\\
  J.~Ambjorn, S.~Arianos, J.~A.~Gesser, S.~Kawamoto,
  Phys.\ Lett.\  {\bf B599 } (2004)  306-312.
  [hep-th/0406108].\\
  J.~Ambjorn, J.~A.~Gesser,
  Phys.\ Lett.\  {\bf B659 } (2008)  718-722.
  [arXiv:0707.3431 [hep-th]].
  Phys.\ Lett.\  {\bf B653 } (2007)  439-444.
  [arXiv:0706.3231 [hep-th]].\\
  J.~A.~Gesser,
  [arXiv:1010.5006 [hep-th]].\\
  M.~R.~Atkin, J.~F.~Wheater,
  JHEP {\bf 1102 } (2011)  084.
  [arXiv:1011.5989 [hep-th]].

\bibitem{genus1}
  A.~Gupta, S.~P.~Trivedi, M.~B.~Wise,
  Nucl.\ Phys.\  {\bf B340 } (1990)  475-490.\\
  M.~Bershadsky, I.~R.~Klebanov,
  Phys.\ Rev.\ Lett.\  {\bf 65 } (1990)  3088-3091.
 

\bibitem{kawaigenus}
  H.~Kawai, N.~Tsuda, T.~Yukawa,
  Phys.\ Lett.\  {\bf B351 } (1995)  162-168.
  [hep-th/9503052].
  Nucl.\ Phys.\ Proc.\ Suppl.\  {\bf 47 } (1996)  653-656.
  [hep-lat/9512014].
  H.~Kawai, N.~Tsuda, T.~Yukawa,
  Nucl.\ Phys.\ Proc.\ Suppl.\  {\bf 53 } (1997)  777-779.
  [hep-lat/9609002].

\bibitem{DDK}
  F.~David,
  Mod.\ Phys.\ Lett.\  {\bf A3 } (1988)  1651.\\
  J.~Distler, H.~Kawai,
  Nucl.\ Phys.\  {\bf B321 } (1989)  509.
  

\bibitem{kk}
  N.~Kawamoto, V.~A.~Kazakov, Y.~Saeki, Y.~Watabiki,
  Phys.\ Rev.\ Lett.\  {\bf 68 } (1992)  2113-2116.
  

\bibitem{fractal}
  H.~Kawai, N.~Kawamoto, T.~Mogami, Y.~Watabiki,
  Phys.\ Lett.\  {\bf B306 } (1993)  19-26.
  [hep-th/9302133].\\
  J.~Ambjorn, Y.~Watabiki,
  Nucl.\ Phys.\  {\bf B445 } (1995)  129-144.
  [hep-th/9501049].

\bibitem{fractal1}
  J.~Ambjorn, J.~Jurkiewicz, Y.~Watabiki,
  Nucl.\ Phys.\  {\bf B454 } (1995)  313-342.
  [hep-lat/9507014].


\bibitem{conical}
 P.\ Menotti, P.\ Peirano, Nucl.Phys. {\bf B473} (1996) 426;
Phys.Lett. {\bf B353} (1995) 444. 

\bibitem{ID}
  C.~Itzykson, J.~M.~Drouffe,
  Cambridge, UK: Univ. Pr. (1989) 405-810.

\bibitem{abbl}
J.~Ambjorn, J.~Barkley, T.~Budd, R.~Loll,
arXiv:1110.3998.



\bibitem{book}
  J.~Ambjorn, B.~Durhuus, T.~Jonsson,
  Cambridge, UK: Univ. Pr., 1997. (Cambridge Monographs in 
Mathematical Physics). 363 p.
  
\bibitem{distler}
  J.~Distler,
  Nucl.\ Phys.\  {\bf B342 } (1990)  523-538.
  


\bibitem{kkm}
  V.~A.~Kazakov, A.~A.~Migdal, I.~K.~Kostov,
  Phys.\ Lett.\  {\bf B157 } (1985)  295-300.
  D.~V.~Boulatov, V.~A.~Kazakov, I.~K.~Kostov, A.~A.~Migdal,
  Nucl.\ Phys.\  {\bf B275 } (1986)  641.
  
\bibitem{c-2fractal}
  J.~Ambjorn, K.~N.~Anagnostopoulos, T.~Ichihara, L.~Jensen, 
N.~Kawamoto, Y.~Watabiki, K.~Yotsuji,
  Phys.\ Lett.\  {\bf B397 } (1997)  177-184.
  [hep-lat/9611032];
  Nucl.\ Phys.\  {\bf B511 } (1998)  673-710.
  [hep-lat/9706009].
  Nucl.\ Phys.\ Proc.\ Suppl.\  {\bf 63 } (1998)  748-750.
  [hep-lat/9709063].

\bibitem{knuth2005}
 D. E. Knuth, 
 \emph{The art of Computer programming, Volume 4A, Combinatorial Algorithms, Part1}, Upper Saddle River, NJ: Addison-Wesley, 2011.
 
\bibitem{chapuy2011}
 G. Chapuy, Adv. Appl. Math. {\bf 47} (2011) 874-893 [arXiv:1006.5053].

\bibitem{watabiki}
  Y.~Watabiki,
  Prog.\ Theor.\ Phys.\ Suppl.\  {\bf 114 } (1993)  1-17.


\bibitem{fractal2}
  J.~Ambjorn, K.~N.~Anagnostopoulos, T.~Ichihara, L.~Jensen, Y.~Watabiki,
  JHEP {\bf 9811 } (1998)  022,
  [hep-lat/9808027].\\
  J.~Ambjorn, K.~N.~Anagnostopoulos,
  Nucl.\ Phys.\  {\bf B497 } (1997)  445-478,
  [hep-lat/9701006].

\bibitem{carfora}M. Troyanov, Trans. Amer. Math. Soc. {\bf 324} (1991) 793;
arXiv: math/0702666v2 [math.DG].\\
M. Carfora and A. Marzuoli, {\it Quantum Triangulations}, Springer, 
{\it to appear.} 

\bibitem{gu}
X.~Gu, Y.~Wang, S.-T.~Yau, Commun. Inf. Syst. {\bf 3}, 3 (2003), 153-170

\bibitem{farkas}
  H.M.~Farkas, I.~Kra, \emph{Riemann Surfaces}, 
  Graduate Texts in Math. 71, Springer-Verlag, New York, 1980.
 
\bibitem{geer}
	G.~van~der~Geer, math/0605346
 
\end{thebibliography}
\end{document}